\documentclass{imamat}

\usepackage{graphicx}
\newcommand{\vep}{\varepsilon}

\newcommand{\bJ}{{\boldsymbol J}}

\newcommand{\bQ}{{\boldsymbol Q}}
\newcommand{\bR}{{\boldsymbol R}}

\newcommand{\bT}{{\boldsymbol T}}

\newcommand{\bk}{{\boldsymbol k}}
\newcommand{\bq}{{\boldsymbol q}}

\newcommand{\fbf}{{\boldsymbol f}}

\newcommand{\bv}{{\boldsymbol v}}

\newcommand{\bx}{{\boldsymbol x}}
\newcommand{\by}{{\boldsymbol y}}

\newcommand{\bX}{{\boldsymbol X}}

\newcommand{\brho}{{\boldsymbol \rho}}

\newcommand{\bxi}{{\boldsymbol \xi}}

\newcommand{\fbb}{{\boldsymbol b}}

\numberwithin{equation}{section}

\def\<{\mathop{\rm \langle}\nolimits}
\def\div{\mathop{\rm div}\nolimits}
\def\diam{\mathop{\rm diam}\nolimits}
\def\>{\mathop{\rm \rangle}\nolimits}
\begin{document}

\title{Optimizing performance of the deconvolution model reduction for large ODE systems}

\shorttitle{Optimizing performance of the deconvolution model reduction} 
\shortauthorlist{L. L. Barannyk and A. Panchenko} 

%

\author{
\name{Lyudmyla L. Barannyk$^*$}
\address{Department of Mathematics, University of Idaho, Moscow, ID 83844 \email{$^*$Corresponding author: barannyk@uidaho.edu}}
\author{Alexander Panchenko}
\address{Department of Mathematics, Washington State University, Pullman, WA, 99164}}

\author{
\name{Lyudmyla L. Barannyk$^*$}
\address{Department of Mathematics, University of Idaho, Moscow, ID 83844\email{$^*$Corresponding author: barannyk@uidaho.edu}}
\name{Alexander Panchenko}
\address{Department of Mathematics, Washington State University, Pullman, WA, 99164}
}

\maketitle



\bibliographystyle{imamat}

\begin{abstract}
{We  investigate the numerical performance of the regularized deconvolution closure introduced recently by the authors. 
The purpose of the closure is to furnish constitutive equations for Irwing-Kirkwood-Noll procedure, a well known method for deriving continuum balance equations from the Newton's equations of particle dynamics.  A version of this procedure used in the paper relies on spatial averaging developed by Hardy, and independently by Murdoch and Bedeaux.  The constitutive equations for the stress are given as a sum of several operator terms acting on the mesoscale average density and velocity. Each term is a  ``convolution sandwich"  containing the deconvolution operator, a composition or a product operator, and the convolution (averaging) operator. 
Deconvolution is constructed using filtered regularization methods from the theory of ill-posed problems. The purpose of regularization is to ensure numerical stability. The particular technique used for numerical experiments  is truncated singular value decomposition (SVD). 
The accuracy of the constitutive equations depends on several parameters: the choice of the averaging window function, the value of the mesoscale resolution parameter, scale separation, the level of truncation of singular values, and the level of spectral filtering of the averages. We conduct numerical experiments to determine the effect of each parameter on the accuracy and efficiency of the method. Partial error estimates are also obtained. }
{FPU chain, Êparticle chain, ÊIrwing-Kirkwood-Noll procedure, Hardy-Murdoch averaging, upscaling, model reduction, dimension reduction, closure, regularized deconvolution, ill-posed problems,
truncated SVD, spectral filtering}
\\
2010 Math Subject Classification: 35B27, Ê37Kxx, 70F10, 70Hxx, 74Q10, 82C21, 82C22, 74Q15, 65F22, 65L09 Ê

\end{abstract}


\section{Introduction}
Many biological and advanced man-made materials are highly heterogeneous, with internal structure spanning a broad range of spatial scales. Micron-to-nanometer scales seem to be of particular importance for understanding the response of such materials. Traditional continuum models miss important effects at sub-micron scales: micro-instabilities, dispersive energy transport, radiative damping, and presence of phonon gaps (see e.g. \cite{Charlotte_Truskinovsky_2012}). To capture these features, one could use molecular dynamics (MD), but analysis of such models is difficult, and their computational efficiency is poor. For millimeter size samples, MD simulations can only access time scales on the order of $10^{-8}$ sec, which is unsatisfactory for many applications. This makes it useful to look for mesoscopic models having the efficiency of classical continuum and capable of describing smaller-scale effects. A related problem in computational science is the design of fast multiscale algorithms for simulating coarse scale particle dynamics. Another relevant question, perhaps the most fundamental of the three, is to find a general method for generating continuum constitutive equations from the underlying atomistic models.

Continuum balance equations of can be derived by averaging particle dynamics. This approach was originated 
by  \cite{Irving_Kirkwood_1950}, who used ensemble averages to derive hydrodynamics equations directly from Hamiltonian dynamics. Soon after,
 \cite{Noll} improved the mathematical foundation of the method by introducing finite size ``window functions'' instead of delta functions. Several decades later,  \cite{Hardy} and 
\cite{mb}, \cite{mb96}, \cite{mb97}, \cite{murdoch07} independently developed a theory based entirely on space-time averaging. The averages can be defined for all spatial and temporal scales, and at each scale, no matter how small, they satisfy exact continuum balance equations. In that sense, the theory works the same way at all scales. This is a useful property because it ensures seamless integration of different regions in coupled multiscale simulations. Recently, Hardy-Murdoch approach has attracted much attention in the applied mathematics, materials science, and physics communities. It has found use both for solids by \cite{Tadmor_Miller_2011} and fluids far from equilibrium by \cite{Evans_Morriss_2008}. These books also contain a large number of references. A few representative examples are \cite{E_Ren_Vanden-Eijnden_2009} (complex fluids), \cite{Lehoucq_Sears_2011} (peridynamics),  and \cite{Zimmerman_2010} (solids). In the latter work, the authors develop a material (Lagrangian) frame approach as opposed to spatial (Eulerian) formulation of the standard Hardy-Murdoch equations. The recent papers by \cite{Admal_Tadmor_2010, Admal_Tadmor_2011} extend the averaging approach to multibody potentials.

Despite its many attractive features, Hardy-Murdoch averaging does not produce an effective continuum theory.  
The fluxes in their balance equations are functions of particle positions and velocities. This means that trajectories of all particles must be known before fluxes can be calculated. In contrast, constitutive equations of classical continuum express fluxes in terms of density, velocity, deformation, and temperature. To recover the efficiency of continuum description, one needs to formulate constitutive equations for Hardy-Murdoch fluxes, that is,
approximate them by operators acting on the averages. 
These approximations would play the same role as constitutive equations of classical continuum: they would decouple balance equations from the underlying particle dynamics.

Meso-scale continuum theories differ from classical continuum. The two features they exhibit most often are non-locality and scale dependence. For example, spatially non-local solid mechanics theories were proposed by  \cite{Eringen_1976} and  \cite{Kunin_1982}.  \cite{Charlotte_Truskinovsky_2012} introduced  a temporally non-local (history dependent) model of one-dimensional lattices with linear nearest neighbor interactions. Another non-local and scale dependent approach is peridynamics proposed by  \cite{Silling_2000}. In the past 12 years peridynamics has gained popularity and has been the subject of many works in applied mathematics, notably by
Du, Gunzburger, Lehoucq, Silling and many others (see the review papers by \cite{Silling_Lehoucq_2010, Lehoucq_Sears_2011, Du_Gunzburger_Lehoucq_Zhou_2012} and references therein). The original theory is phenomenological, so the constitutive equations are postulated or fitted from experiments. Scale dependence is introduced by using a fixed size window function (called horizon by Silling).  The latter feature is similar to Hardy-Murdoch procedure, so one expects that peridynamic equations can be derived by averaging atomistic models. This was recently done by  \cite{Lehoucq_Sears_2011}. They provided an exact description of the peridynamical force state in terms of the particle positions and velocities.

A method for generating non-local and scale-dependent constitutive equations for Hardy-Murdoch averages was introduced in \cite{Panchenko_Barannyk_Gilbert_2011} and tested numerically in \cite{Panchenko_Barannyk_Cooper}. Applications to discrete models of fluid flow were studied in \cite{Tartakovsky_Panchenko_Ferris_2011}, \cite{Panchenko_Tartakovsky}. Recently, similar methods have been successfully used in large eddy simulation of turbulence by \cite{Layton_2006, Kim_Rebholz_Fried_2012, Layton_Rebholz_2012}. 
An outline of the method is as follows. The fine scale model is a system of Newton's ODEs for $N$ particles. A typical interparticle distance is characterized by a small parameter $\vep=N^{-1/d}$, where $d$ is the dimension of the physical space. The spatial averages are defined using a window function depending on the mesoscale resolution parameter $\eta\gg \vep$.
\begin{quote}
\noindent
(i) Rewrite non-linear averages as (linear) convolutions of the window function and certain functions of particle trajectories. The number of these micro-scale dynamical functions is small and equals  the number of continuum quantities in the theory.  A proper choice of the window function
insures invertibility of the convolution. Thus the above micro-scale dynamical functions are {\it recoverable}, at least in principle.\newline

\vspace*{-0.2cm}
\noindent
(ii) The difficulty here is that deconvolution problem is unstable (ill-posed), so small errors in the averages may produce large errors in the recovered functions. Instability can be overcome by using regularization methods. Regularization produces stable approximations of the micro-scale quantities in terms of the averages.\newline

\vspace*{-0.2cm}
\noindent
(iii)  In the simplest case of nearly isothermal dynamics, the energy balance is assumed to be trivial. The only flux that requires closure is the stress in the momentum balance equation. The observable averages in this case are density $\overline{\rho}^\eta$ and linear momentum $\overline{\rho}^\eta \overline{\bv}^\eta$. The corresponding recoverable functions are, respectively,  Jacobian $J$ of the inverse fine scale deformation map, and the product $J\tilde \bv$ of this Jacobian and a fine scale velocity interpolant $\tilde \bv$. More details on the definition of these quantities can be found in 
\cite{Panchenko_Barannyk_Gilbert_2011, Panchenko_Barannyk_Cooper} and in the brief description provided in Section \ref{governing_equations}. \newline

\vspace*{-0.2cm}
\noindent
(iv) Next, we take advantage of a remarkable property of Hardy-Murdoch averages: the exact stress can be shown to depend only on the recoverable functions $J$ and $\tilde\bv$. Thus we can substitute their deconvolution approximations into the exact flux equations and obtain (approximate) constitutive equations. This produces a meso-scale system in closed form that is completely decoupled from the underlying MD.
\end{quote}
The constitutive equations express stress in terms of the average density and average velocity. Fine scale information is also incorporated, but the nature of this information is such that solving the ODEs is no longer required. Typically we would use the initial conditions and the equation
for the interatomic potential, but nothing else. 
The main structural unit of the stress equations is a ``convolution sandwich"
\begin{equation}
\label{sandwich}
R_\eta S Q_\eta, 
\end{equation}
where $R_\eta$ is the convolution (averaging) operator, $Q_\eta$ is the deconvolution operator (this is an approximate regularized inverse of $R_\eta$), and $S$ is a nonlinear composition or a product type operator. The complete stress is a sum of the terms of the form (\ref{sandwich}) where 
$Q_\eta[\overline{\rho}^\eta]$ and $Q_\eta[\overline{\rho}^\eta \overline{\bv}^\eta]$ may be present.

In this paper we study algorithmic realization of the regularized deconvolution closure.  Since the problem at hand is nonlinear, the error estimates that we provide are not likely to be tight, so we also study the error evolution numerically. The objective is to understand the dependence of the error on various parameters of the method.  
Specifically, we study the effects of a choice of a window function $\psi$, resolution parameter $\eta$, scale separation, truncation level in SVD and filtering of spectral coefficients.
%
The numerical experiments are conducted as follows. We solve the system of particle dynamics ODEs and compute all  relevant averages. Next, we apply regularized deconvolution to computed average density and momentum to obtain approximations of $J$ and $\tilde\bv$.
These approximations are used to calculate the approximate stress that is then compared to the directly computed (``exact") stress.
For simulation of particle dynamics, we use two sets of the test initial conditions. The initial positions for both of them are uniformly spaced. The initial velocities in the first test case are a discretized low frequency sine function. In the second case, the initial velocities are prescribed by a truncated $4$th polynomial imitating a  Gaussian. The corresponding Fourier spectrum is full (unlike the first test case) and decays at a polynomial rate.

A discretization of the integral convolution equation is a linear system. As a consequence of ill-posedness, the matrix of this system is ill-conditioned. To regularize the discrete problem, we use truncated singular value decomposition (SVD): the exact solution is approximated by its projection onto the subspace spanned by singular vectors corresponding to larger singular values. The number of the retained singular values plays the role of regularization parameter.
Since the convolution kernel is dynamics-independent, SVD should be computed only once. This can be done prior to running time-dependent simulations. The same kernel may be used with different dynamical systems, and thus the cost of computing SVD can be excluded from the operation count of the dynamical simulation. The algorithms for computing SVD are readily available from the standard linear algebra packages such as LAPACK \cite{LAPACK_Users_Guide}. All of the above makes SVD-based regularizations convenient and easy to implement. However, using SVD has several drawbacks. Many standard SVD algorithms are iterative, and smaller singular are computed less accurately. The same is true for computation of the associated singular vectors. In addition, the performance of the standard algorithms including LAPACK  becomes worse as the condition number of the matrix increases. 
For numerical experiments performed in this work, advantages of using the truncated SVD  seem to outweigh the disadvantages.

In the continuum case, degree of ill-posedness depends on the rate of decay of the singular values of the kernel (see \cite{Kirsch_1996}). This rate in turn depends on the smoothness of the kernel (see \cite{Hansen_1987}). It is then reasonable to expect that the rate of decay of singular values plays a similar role in the discrete setting. However, the overall error, that is, the error in computing the approximate stress (\ref{sandwich}), does follow this pattern.  The Gaussian kernel, which has the fastest decay rate among the considered window functions,  corresponds to a smaller overall error than less smooth piecewise linear kernels. We consider six different window functions: characteristic function of an interval, piecewise linear trapezoid function and piecewise linear  triangular function (finite element function), truncated $2$nd and $4$th order polynomials and Gaussian function. Among these functions, the Gaussian and truncated $4$th order polynomials give the least error. Occasionally the truncated $4$th order polynomial produces slightly smaller error than the Gaussian but using the Gaussian requires a significantly smaller number of singular values to achieve comparable accuracy.
 
An important aspect of the method's performance is dependence of the error on scale separation characterized by, for example, the product of $\eta $ and 
$N$. Ideally, the error should decrease as scale separation increases. We test the method in two regimes: one corresponds to increasing $\eta$ keeping $N$ fixed, and the other consists of increasing $N$ while $\eta$ is fixed.  These regimes are not equivalent. The first corresponds to taking a fixed fine scale dynamical system and increasing the size of the averaging window. The second regime is different: we
keep the size of the window fixed, but the underlying particle systems vary as follows. We increase $N$ while keeping the total mass fixed, and rescale the forces so that the total energy $E(N)$ is bounded independent of $N$.
The initial conditions in this regime are given by increasingly finer discretizations of the same continuum functions.  Existence of the uniform in $N$ bound on energy does not imply monotonic dependence on $N$, and thus it is unlikely that the averages of different particle systems depend on $N$ monotonically.

In the first regime, as $\eta$ increases, the density and linear momentum tend to get smoother and converge to zero as $\eta\to\infty$ (see equations (\ref{av3}) and (\ref{av4})). The exact stress depends on $\eta$ similarly, but the dependence of the approximate stress on $\eta$ is more complicated. To form the approximate stress, we replace the exact recoverable functions by their deconvoluted approximations. The latter depend on $\eta$, and thus the nonlinear functions of the recoverable functions  appearing  in (\ref{sandwich}) become $\eta$-dependent. Finally, to produce the approximate stress, these nonlinear functions are averaged with the $\eta$-dependent window function. 
The quality of the approximation is thus determined by two competing tendencies.
On the one hand, for larger $\eta$, one expects worsening of the reconstruction $Q_\eta$ because the same amount of fine scale information is harder to extract from increasingly smoother and smaller data. On the other hand, the averaging $R_\eta$ in (\ref{sandwich}) decreases the total error by damping the high frequency modes. 
The error in the deconvolution $Q_\eta$  is primarily at high frequencies. Application of the nonlinear dispersive operator $S$ may cause propagation of the error into low frequency harmonics. This effect may be difficult to counteract by applying  $R_\eta$, but in simulations  the overall relative error tends to decrease with increasing $\eta$, even though the deconvolution error increases with $\eta$. This seems to be a remarkable feature of the deconvolution closure: the stress is computed more accurately despite the fact that the recoverable functions are approximated more poorly. This shows that the constitutive equations introduced in \cite{Panchenko_Barannyk_Gilbert_2011, Panchenko_Barannyk_Cooper} become more accurate at larger scales.
We also confirm this by comparing Fourier spectra of the exact recoverable functions and stresses to the spectra of their approximations. For both test cases, the truncated SVD approximation captures low frequency harmonics but looses high frequencies. The Fourier spectra of the stress are dominated by  low frequency harmonics and they are approximated  well. It seems that dispersion due to the presence of $S$ in (\ref{sandwich}) is relatively weak.

In the second regime, we simulate evolution the error as $N$ increases. We observe that for certain times the error is not monotonic in $N$. This deviation from monotonicity seems to be related to the behavior of the computed total energy of the system. While the exact total energy is conserved, the computed energy oscillates   in time in a nearly independent of $N$ manner.  The oscillations are between the initial value (the exact energy of the system), and a slightly larger (by at most $0.05\%$) value. At times when the computed energy is close to the initial value, the error decreases as $N$ increases. At other times, the dependence of the error on $N$ is not monotonic. It appears that enforcing conservation of energy numerically, for example, by using energy-preserving Runge-Kutta methods (see \cite{Celledoni_2009}), should ensure monotonic decrease of the error with $N$.

The classical theory of ill-posed problems \cite{Kirsch_1996, Tikhonov_Arsenin_1987, Morozov_1984} deals mostly with operators between Hilbert spaces, because for integral operators, the Hilbert space structure is required for existence of singular value decomposition. In the discrete setting, one can use SVD with any $p$-norm, including the $\infty$-norm. In the paper, we provide such error estimates for $p\in [1, \infty]$. The values of $p$ for the right hand side and the solution may be different.  We also use the $\infty$-norm for computational assessment of the quality of approximation.

The organization of the paper is as follows. The background concerning the fine scale problem and spatial averaging is briefly described in Section \ref{governing_equations}.
Section \ref{SVD} contains the necessary facts about filtered regularization methods in the discrete setting. In Section \ref{window_function} we study various window functions used to set-up averaging and the effect of the choice of a window function on the quality of the stress approximation. 
In Section \ref{resolution_parameter} we analyze  the effect of the resolution parameter on the reconstruction of microscopic quantities of interest and eventual stress approximation.  
In Section \ref{spectral_evolution} we study the Fourier spectra of the reconstruction error and the overall error in the stress.
%
%
The effect of the scale separation is considered in Section \ref{scale_separation}. 
Error estimates for filtered regularization methods for $p\in [1, \infty]$ are provided in Section \ref{error_estimates}. Finally, in Section \ref{conclusions} we give conclusions. In appendices, we define various window functions used in the numerical experiments and provide the formula for the Lennard-Jones potential.

\section{Governing equations} \label{governing_equations}

\subsection{Microscale equations}
The object of our study is a system of $N$ particles that move according to Newton's law of motion, where $N\gg 1$. Particle positions and velocities are denoted by ${\bq}_i(t)$ and ${\bv}_i(t)$, $i=1,\ldots, N$, respectively. For simplicity, we assume that all particles have the same mass $M/N$ where $M$ is the total mass of the system.  The equations of motion are
\begin{eqnarray}
\dot\bq_i&=&\bv_i,\label{one}\\
 \frac{M}{N}\dot\bv_i&=&\fbf_i+\fbf_i^{(ext)},\label{two}
\end{eqnarray}
with initial conditions
\begin{equation}
\label{inits}
\bq_i(0)=\bq_i^{0}, \hspace*{1.0cm} \bv_i(0)=\bv^0_i.
\end{equation}
Here $\fbf_i^{(ext)}$ is an external force (gravity) and $\fbf_i=\sum_j \fbf_{ij}$,  
where $\fbf_{ij}$ is an  interaction force exerted on a particle  $i$ by a particle $j$. 
These  forces are generated by a pair potential $U(|{\bq}_i-{\bq}_j|)$. In numerical experiments we use the classical Lennard-Jones potential (see equation (\ref{LJ_def})). 

To study the behavior of the system for large $N$, it is convenient to introduce a small parameter $\varepsilon=N^{-1/d}$, where $d=1,2,3$ is the dimension of the physical space, and a mesoscopic resolution parameter $\eta$ satisfying $0<\eta<1$ and  $\varepsilon \ll \eta$.
Using these parameters we  define the microscopic length scale $\varepsilon L$ and the mesoscopic length scale $\eta L$, where  $L=\diam(\Omega)$ and  $\Omega$ is the computational domain. 
To make the total energy of the system bounded independent of $N$, we scale interparticle forces by $\varepsilon$ as in \cite{Panchenko_Barannyk_Cooper}.

\subsection{Mesoscale averaging and dynamics} \label{mesoscale_averaging}

The first step of model reduction in  \cite{Panchenko_Barannyk_Gilbert_2011, Panchenko_Barannyk_Cooper} is to use space-time averaging pioneered by  \cite{Noll},  \cite{Hardy} and   \cite{mb, murdoch07}. In this section, we briefly describe the averaging approach to make the exposition self-contained. For simplicity we consider only spatial averaging and refer to \cite{Noll}, \cite{Hardy} and  \cite{mb, murdoch07}  for more details.

To set up spatial averaging, choose a fast decreasing window function $\psi(x)\geq 0$ such  that $\int_{-\infty}^\infty\psi(x)=1$. We assume that $\psi$ is 
continuous and differentiable almost everywhere on the interior of its support. 
After scaling by scaling by $\eta$ , the function
$
\psi_{\eta}(x)=\frac 1\eta \psi\left(\frac x\eta\right)
$
is used to define the averages of microscopic quantities. For example, the approximate number of particles within the distance $\eta L$ of a point $\bx$ is
\[
n^\eta(t, \bx)=\sum_{i=1}^N \psi_\eta(\bx-\bq_i(t)).
\]
Similarly, one defines the average density and linear momentum:
%
%
\begin{equation}
\label{density}
\overline{\rho}^\eta(t, \bx)=\frac{M}{N}\sum_{i=1}^N \psi_\eta(\bx-\bq_i(t)),
\end{equation}
\begin{equation}
\label{mom}
\overline{\rho}^\eta \overline{\bv}^\eta(t, \bx)=\frac{M}{N}\sum \bv_i(t)\psi_\eta(\bx-\bq_i(t)).
\end{equation}

Differentiating equations (\ref{density}), (\ref{mom}) with respect to $t$, and using the ODEs (\ref{one}), (\ref{two}) yields
exact balance equations for $\overline{\rho}^\eta$ and $\overline{\rho}^\eta\overline{\bv}^\eta$ (see \cite{Hardy, mb}):
\begin{equation}
\label{mass-balance}
\partial_t \overline{\rho}^\eta+\div (\rho^\eta\overline{\bv}^\eta)=0,
\end{equation}
\begin{equation}
\label{m-balance}
\partial_t(\overline{\rho}^\eta\overline{\bv}^\eta)+\div\left(\overline{\rho}^\eta\overline{\bv}^\eta\otimes
\overline{\bv}^\eta\right) - \div\bT^\eta=0.
\end{equation}
Here the system is assumed to be isolated ($\fbf_i^{(ext)}=0$); $\bT^\eta$ is the total stress written as
$\bT^\eta=\bT^\eta_{(c)}+\bT^\eta_{(int)}$ (see \cite{murdoch07}), where
\begin{equation}
\label{m-stress-c}
\bT^\eta_{(c)}(t, \bx)=-\sum m_i(\bv_i-\overline{\bv}^\eta(t, \bx, ))\otimes (\bv_i-\overline{\bv}^\eta(\bx, t))\psi_\eta(\bx-\bq_i)
\end{equation}
is the {\it convective stress}, and
\begin{equation}
\label{m-stress-int}
\bT^\eta(t, \bx)_{(int)}=
\sum_{(i, j)}\fbf_{ij}\otimes (\bq_j-\bq_i)\int_0^1 \psi_\eta\left(s(\bx-\bq_j)+(1-s)(\bx-\bq_i)\right)ds
\end{equation}
is the {\it interaction stress}. The summation in (\ref{m-stress-int}) is over all pairs of particles $(i, j)$ that interact with each other. 
A similar energy balance equation can also be written if needed (see \cite{Hardy, mb}).
We refer to (\ref{mass-balance}), (\ref{m-balance}) as the {\it meso system}. The unknowns $\overline{\rho}^\eta$ and $\overline{v}^\eta$ are at the mesoscale whereas the stress $\bT^\eta$ depends on microscopic variables $\bq_i$ and $\bv_i$. Hence, the meso system is not closed because one
must know the trajectories of all particles before $\bT^\eta$ can be determined.
Therefore, solving the exact meso system is no more efficient than solving the original ODEs (\ref{one}), (\ref{inits}).  
\subsection{Integral approximations and closure}
To improve efficiency, we use the closure method introduced in \cite{Panchenko_Barannyk_Gilbert_2011, Panchenko_Barannyk_Cooper}.  In this section, we outline the method for convenience of the reader.

The objective is to find an approximation of $\bT^\eta$ in terms of $\overline{\rho}^\eta$ and $\overline{\rho}^\eta \overline{\bv}^\eta$.
First, observe that the discrete sums in  (\ref{density}), (\ref{mom}) can be approximated by the convolution integrals 
\begin{equation} \label{conv_oper}
R_\eta[f](\bx)=\int \psi_\eta(\bx-\by) f(\by) d\by \approx \overline{f}^\eta.
\end{equation}
Here $R_\eta$ is a convolution operator, $f$ is a microscopic quantity to be recovered, $\overline{f}^\eta$ is the known mesoscale average. The integral approximations of the average density and average momentum are as follows.
\begin{eqnarray}
\label{int-density}
\overline{\rho}^\eta(t, \bx)&  =  &
\frac{M}{N}\sum_{i=1}^N \psi_\eta(\bx-\bq_i(t))
\approx  \frac{M}{|\Omega|}\int_\Omega \psi_\eta \left(\bx-\tilde\bq(t, \bX)\right) d\bX \\
& \approx  &\frac{M}{|\Omega|}\int_\Omega \psi_\eta (\bx-\by) J(t, \by) d\by =\frac{M}{|\Omega|} R_\eta[J](\bx),\nonumber
\end{eqnarray}
\begin{equation}
\label{int-mom}
\overline{\rho}^\eta \overline{\bv}^\eta(t, \bx)= \frac{M}{N}\sum_{i=1}^N \bv_i(t)\psi_\eta(\bx-\bq_i(t))\\
 \approx  \frac{M}{|\Omega|}\int_\Omega \psi_\eta (\bx-\by) \tilde{\bv}(t, \by) J(t, \by) d\by=  \frac{M}{|\Omega|} R_\eta[\tilde{\bv} J](\bx).
\end{equation}
Here $M$ is the mass of the system, and $|\Omega|$ is  the volume (Lebesgue measure) of the domain $\Omega$ occupied by the system. 
The functions $\tilde{\bq}(t, \bX)$,  $\tilde{\bv}(t, \tilde{\bq})$ are suitable interpolants of particle positions and 
velocities. The Jacobian
\begin{equation}\label{Jac}
J=|\det \nabla \tilde{\bq}^{-1}|
\end{equation}
describes local volume changes. 

Integral approximations of the stress are obtained similarly (see \cite{Panchenko_Barannyk_Cooper}).
To approximate the interaction stress $\bT^\eta_{(int)}$ we write $\bq_i=\tilde\bq(t, \bX_1), \bq_j=\tilde\bq(t, \bX_2)$, and approximate the double sum in
(\ref{m-stress-int}) by the double integral with respect to $\bX_1, \bX_2$. 
After several changes of variables of integration we arrive at 
\begin{equation}
\label{c2-int}
\bT^\eta_{(int)}(t, \bx)=
\frac{1}{|\Omega|^2}\int \psi_\eta(\bx-\bR) 
\left(\int U^\prime(|\brho|)\frac{\brho\otimes \brho}{|\brho|}
J(t, \bR+\frac{\vep}{2}\brho)J(t, \bR-\frac{\vep}{2}\brho) d\brho
\right)\; d \bR.
\end{equation}
Equation (\ref{c2-int}) can be written compactly as
\begin{equation}
\label{compact-int-stress}
\bT^\eta_{(int)}=R_\eta S_1 [J],
\end{equation}
where $R_\eta$ is the convolution operator, and the operator $S_1$ is a composition of
shifted multiplication $J (t, \bR)\to
J(t, \bR+\frac{\vep}{2}\rho) J(t, \bR-\frac{\vep}{2}\brho)$ and 
 $\brho$-integration with the weight $|\Omega|^{-2}|\brho|^{-1}\brho\otimes\brho ~U^\prime(|\brho|)$.

Convective stress $\bT^\eta_{(c)}$ can be approximated similarly:
\begin{eqnarray}
\label{conv-int}
\bT^\eta_{(c)}(t, \bx)
& \approx  &-
\frac{M}{|\Omega|}\int_\Omega \psi_\eta(\bx-\by)
(\tilde\bv(t, \by)-\overline{\bv}^\eta(t, \bx, ))\otimes
(\tilde\bv(t, \by)-\overline{\bv}^\eta(\bx, t))J(t, \by)
d\by\\
& = & 
R_\eta S_2[J, \tilde\bv ], \label{compact-conv-stress}
\end{eqnarray}
where the operator $S_2$ is a multiplication of $-
\frac{M}{|\Omega|}J$ by the diadic product of $\tilde\bv(t, \by)-\overline{\bv}^\eta(t, \bx)$ with itself.

Equations (\ref{c2-int} and (\ref{conv-int}) show that stress can be represented as an operator acting on the two fine scale functions $J$ and $\tilde \bv$. These functions can be approximately recovered by inverting the convolution operator  $R_\eta$ (see (\ref{int-density}) and (\ref{int-mom})).
The difficulty here is that straightforward inversion is unstable. Indeed, $R_\eta$ is compact, hence it is not surjective. This implies that $R_\eta^{-1}$ is unbounded, and thus small perturbations of the right hand side may produce large perturbation in the computed solution. On the discrete level, the matrix discretization of $R_\eta$ will have a large condition number. 
Therefore, the deconvolution problem (reconstructing $f$ from the knowledge of $R_\eta[f]$) is ill-posed. Such problems can be approximately solved using regularization (see, e. g.  \cite{Kirsch_1996,  Morozov_1984, Tikhonov_Arsenin_1987}). Regularization consists in approximating the exact unstable problem by a parameter-dependent stable problem. The solution operator $Q_\eta$ of this problem provides an approximation to the exact inverse operator. Computational implementation of regularized deconvolution is described in Section \ref{SVD} below.

The meso system  (\ref{mass-balance}), (\ref{m-balance}) is closed by inserting the approximations
\begin{equation}
\label{closure-app} 
J \approx M^{-1}|\Omega|Q_\eta[\overline{\rho}^\eta], \;\;\;\;\; \tilde \bv\approx \frac{Q_\eta[\overline{\rho}^\eta\overline{\bv}^\eta]}{Q_\eta[\overline{\rho}^\eta]}
\end{equation}
into (\ref{compact-int-stress}), (\ref{compact-conv-stress}).
The resulting operator structure is a sum of terms of the form (\ref{sandwich}), acting on 
$\overline{\rho}^\eta$ and $\overline{\rho}^\eta\overline{\bv}^\eta$.

Computational realization of the closure equations is obtained by approximating microscopic positions using (\ref{Jac}). In one dimension, this is particularly simple, since the Jacobian coincides with the gradient. Approximating it by standard finite difference discretizations, we can generate approximate positions. Then approximate positions and velocities are inserted in (\ref{m-stress-c}) and (\ref{m-stress-int}).


\subsection{Test cases} \label{testICs}
\begin{figure}[h] \centerline{
\includegraphics[width=4in,angle=0]{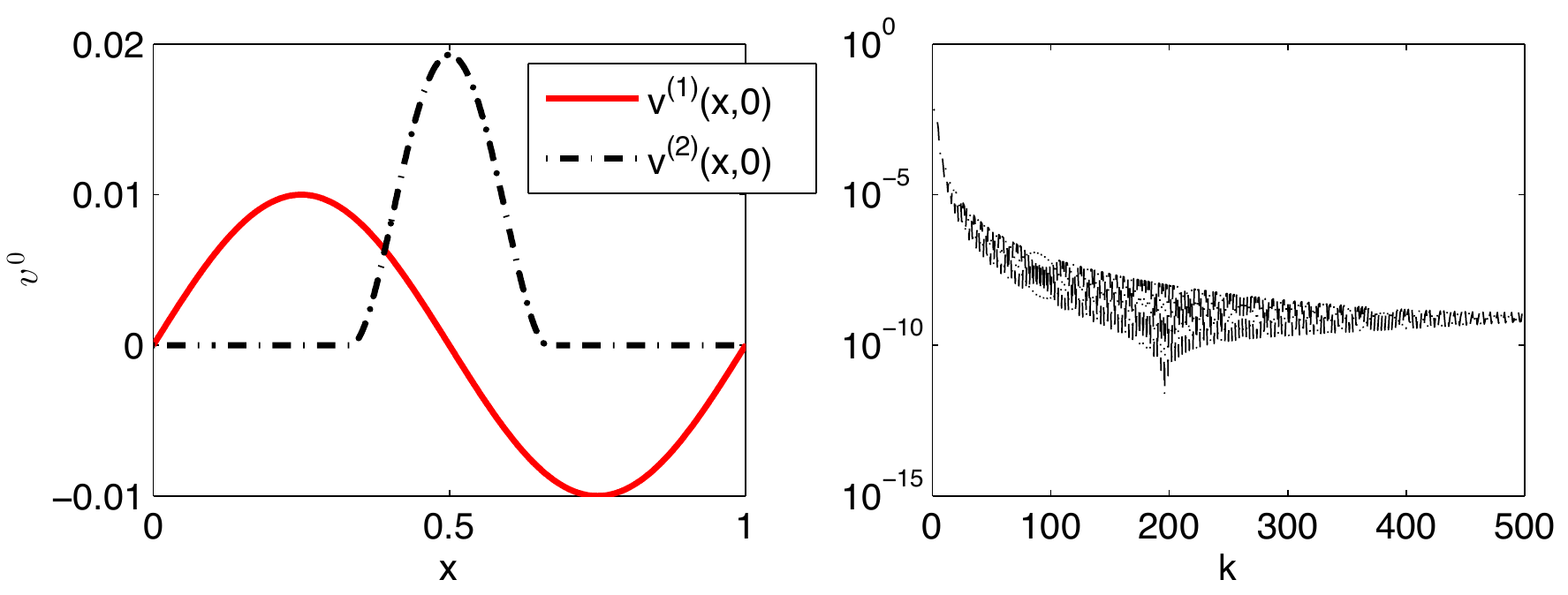}
} \caption{Initial velocities and their Fourier spectrum. Left panel shows initial velocities. Red solid curve is for the first test initial velocity $v^{(1)}(x,0)$ defined in (\ref{ICsine}), black curve --  second test initial velocity $v^{(2)}(x,0)$ given in (\ref{IC4th}). Right panel has discrete Fourier transform of  $v^{(2)}(x,0)$. 
}
\label{W1IC}
\end{figure}

 For computational testing we use Lennard-Jones potential defined in (\ref{LJ_def}) and two sets of the initial conditions. The problem is assumed to be periodic with period $L$. The initial positions in both cases are equally spaced with 
\begin{equation}\label{ICpos}
q_j^{0}=\left(j-\frac 12\right)\Delta x, \quad \mbox{where} \quad \Delta x=\frac LN, \quad j=1,\ldots,N.
\end{equation}
{\bf First test case}. The initial velocity  is a one mode sine function
\begin{equation} \label{ICsine}
v^{(1)}(x,0)=10^{-2} \sin\frac{2\pi x}{L}, \quad 0 \leq x\leq L.
\end{equation}
{\bf Second test case}.   The initial velocity  is a 
continuous function that is a truncated $4$th order polynomial on $[\frac L3, \frac{2L}{3}]$ with double roots at $x=\frac 13$ and $x=\frac 23$ and zero otherwise: 
\begin{equation}\label{IC4th}
 v^{(2)}(x,0)=
 \left\{
 \begin{array}{l}
25(x-\frac 13)^2(x-\frac 23)^2, \quad \mbox{if} \quad \frac{L}{3} \leq x\leq \frac{2L}{3}, \\[5pt]
0, \quad \mbox{otherwise.} 
 \end{array}
 \right.
\end{equation} 
The initial velocities are shown in the left panel of Fig. \ref{W1IC}.
%
The main difference between these initial velocities can be seen by looking at their Fourier spectra.  
The first initial velocity has only one low frequency Fourier  mode while the second initial velocity has all Fourier modes, shown in the right panel of Fig. \ref{W1IC},  present that level off at $10^{-9}$ for $N=1000$. As $N$ increases, the tail of the Fourier spectrum levels off at a slightly lower value (not shown here). For example, for $N=10,\!000$, the tail is at $10^{-12}$. Therefore, we expect to see more complicated nonlinear dynamics and a stronger effect of high frequencies in the second case.

The 1D version of system of ODEs (\ref{one}), (\ref{two}), (\ref{inits}) with the initial conditions defined in (\ref{ICpos}), (\ref{ICsine}) and (\ref{IC4th}) is solved using the Velocity Verlet method until $t=1$ for various $N$.


\section{Filtered regularization methods} \label{SVD}

On a discrete level,  equations (\ref{int-density}), (\ref{int-mom}) 
reduce to a linear system 
%
%
\begin{equation}
\label{e1}
A^\eta \bx=\fbb,
\end{equation}
where $\fbb$ is  a known average quantity such as $\frac{|\Omega|}{M}\overline{\rho}^\eta$ and $\frac{|\Omega|}{M}\overline{\rho}^\eta \overline{\bv}^\eta$ and $\bx$ is either $J$ or $\tilde{\bv} J$. To produce $\fbb$, averages are discretized on a coarse mesh with $B$ nodes, $B\sim 1/\eta$ and $B\ll N$, and the solutions are rendered on a finer mesh with $N^\prime\geq B$ nodes.
In practice, $N^\prime$  can vary between $B$ and $N$. In our numerical experiments we use $N'=N$. 

The matrix $A^\eta$, obtained by discretizing the  kernel $\psi_\eta$, has dimensions $B\times N^\prime$ and $r={\rm rank}(A^\eta)\leq B$.  There are two difficulties associated with solving (\ref{e1}): (i) the condition number of  $A^\eta$ is  large because of the ill-posedness; and  (ii) the system is underdetermined and has multiple solutions.

Performing singular value decomposition (SVD) of $A^\eta$ we obtain $r$ non-zero singular values $\sigma_j$ and left and right singular vectors $\bxi_j$ and  $\hat\bxi_j$,  satisfying
\begin{equation}
\label{e2}
A^\eta\hat \bxi_j=\sigma_j \bxi_j, \;\;\;\; (A^\eta)^T \bxi_j=\sigma_j \hat\bxi_j,\;\;\;\;\;j=1, 2,\ldots, r.
\end{equation}
The singular vectors $\bxi_j$,  $\hat\bxi_j$ have length $B$ and $N$, respectively, and are orthonormal in their respective spaces.
In this paper we work only with matrices $A^\eta$ satisfying
\begin{equation}
\label{A-spectrum}
\sigma_j\in (0, 1].
\end{equation}
This assumption holds for discretizations of all convolution kernels under consideration. 

Expanding the right hand side
\begin{equation} \label{b_expansion}
\fbb=\sum_{j=1}^r b_j \bxi_j,
\end{equation}
we can write the minimal norm solution of (\ref{e1}) 
\begin{equation}
\label{e3}
\bx=\sum_{j=1}^r x_j\hat\bxi_j, \qquad x_j=\frac{b_j}{\sigma_j}, \;\;\;\;\; j=1, 2,\ldots, r.
\end{equation}
This solution  is  orthogonal to the null-space of $A^\eta$.
When the condition number of $A^\eta$ is large, the solution is highly unstable. To stabilize the computation, one can use regularization (see, e. g. \cite{Kirsch_1996, Hansen_1987}). Here we limit ourselves to filtered linear regularization methods that 
replace the exact solution (\ref{e3}) with the approximation
\begin{equation}
\label{e6}
\bx^\alpha= \sum_{j=1}^r b_j \frac{\phi(\sigma_j, \alpha)}{\sigma_j}\hat\bxi_j.
\end{equation}
The function $\phi$ is called a filter function. Generally it should satisfy (see \cite{Kirsch_1996})
\begin{enumerate}
\item $|\phi(s, \alpha)|\leq 1$ for all $\alpha>0$ and $s\in(0, 1]$;
\item for every $\alpha>0$ there is $c(\alpha)$ such that $|\phi(s, \alpha)|\leq c(\alpha)s$  for all 
$s\in (0, 1]$;
\item $\lim_{\alpha\to 0}\phi(s, \alpha)=1$ for every $s\in (0, 1]$.
\end{enumerate}
By Theorem 2.6 in \cite{Kirsch_1996}, (\ref{e6}) defines a regularized approximate inverse $R_\alpha$ to $A^T A$ that has the property $\lim_{\alpha\to 0} R_\alpha (A^T A) \bx \to \bx$ in $2$-norm for each $\bx \in {\mathbb R}^N$.

Examples of filtered methods are Tikhonov regularization with
\begin{equation}
\label{e7}
\phi_T^\alpha=\frac{\sigma_j^2}{\sigma_j^2+\alpha},
\end{equation}
truncated SVD with
\begin{equation}
\label{e8}
\phi^\alpha=
\left\{
\begin{array}{cc}
1& \;{\rm if}\;\sigma_j\leq \sigma_\alpha\\
0 & {\rm otherwise},\\
\end{array}
\right.
\end{equation}
and
Landweber iteration with
\begin{equation}
\label{e9}
\phi_L^n=1-(1-\sigma_j^2)^{n+1}.
\end{equation}
In (\ref{e8}), the role for regularization parameter is played by the cut-off value $\sigma_\alpha$, while for Landweber iteration the regularization parameter is the reciprocal of the number $n$ of iterations. 
The SVD approach may be more efficient than iterative methods (see \cite{Hansen_1987})
 because SVD can be pre-computed and used at each time step for different right hand sides.
The algorithmic realizations of (\ref{e7})--(\ref{e9}) and other numerical regularizing techniques are discussed in the book by \cite{Hansen_1987}.

In this paper, we use a  truncated SVD method to  compute the minimum norm solution (\ref{e3}) and we retain only those singular values whose magnitude is greater than $\sigma_\alpha=10^{-13}$. 
In addition 
we also use spectral filtering technique on the right hand side $\fbb$, which is similar to the Fourier filtering used in 
\cite{Krasny_1986_singularity}. 
%
The coefficients $b_j$ that lie below the tolerance value ${\it tol}=10^{-13}$ are set to $0$. 
We find this filtering helpful when both $\sigma_j$ and $b_j$ are close to the machine precision. In that case, both $b_j$, and $x_j$
may carry a large relative error. Better reconstructions are obtained by discarding the corresponding coefficients $x_j=b_j/\sigma_j$ in the computed solution. Varying $tol$ between $10^{-14}$ and $10^{-11}$ does not essentially change approximations to stresses, while setting the filter level at a higher value  increases the error as expected.

To compute SVD of $A^\eta$, we use a standard LAPACK subroutine DGESVD. This routine computes singular values and left and right singular vectors. It should be noted that the accuracy of this subroutine decreases as the singular values decrease. Specifically, each computed singular value $\hat \sigma_i$ differs from true value $\sigma_i$ by at most
\[
|\hat \sigma_i-\sigma_i|\leq p(m,n) \cdot \epsilon \cdot \sigma_1
\]
where $p(m,n)$ is a modestly growing function of $m$ and $n$, $\epsilon=2^{-53}\approx 1.11\cdot 10^{-16}$ is machine epsilon, and $\sigma_1$ is the largest eigenvalue. Thus singular values near $\sigma_1$ are computed to high relative accuracy and small ones may not be. The computed singular vectors are always orthogonal to working precision but their accuracy depends on how close they are to each other, i.e. if $\sigma_i$ is close to nearby singular values. In the present case, small singular values are densely spaced, so the corresponding singular vectors may be inaccurate (see \cite{LAPACK_Users_Guide}). 
 
Jacobi's method (see \cite{Demmel_Gu_Eisenstat_Slapnicar_Veselic_Drmac_1999, Demmel_1997, Demmel_Veselic_1992, Slapnicar_1992})  is another algorithm for computing SVD. Unlike conventional algorithms, this method is capable of high relative accuracy even for smallest singular values. The original version was slower than the standard algorithms but the speed was recently improved by   \cite{Drmac_Veselic_2007_I, Drmac_Veselic_2007_II}. Subroutines DGEJSV and DGESVJ are available as a part of the current version of LAPACK.
It would be interesting to see if using Jacobi's method  improves the accuracy of the stress approximation.

%
%
%
%

%
%



\section{Choice of a window function} \label{window_function}

\begin{figure}[h] \centerline{
\includegraphics[width=3.5in,angle=0]{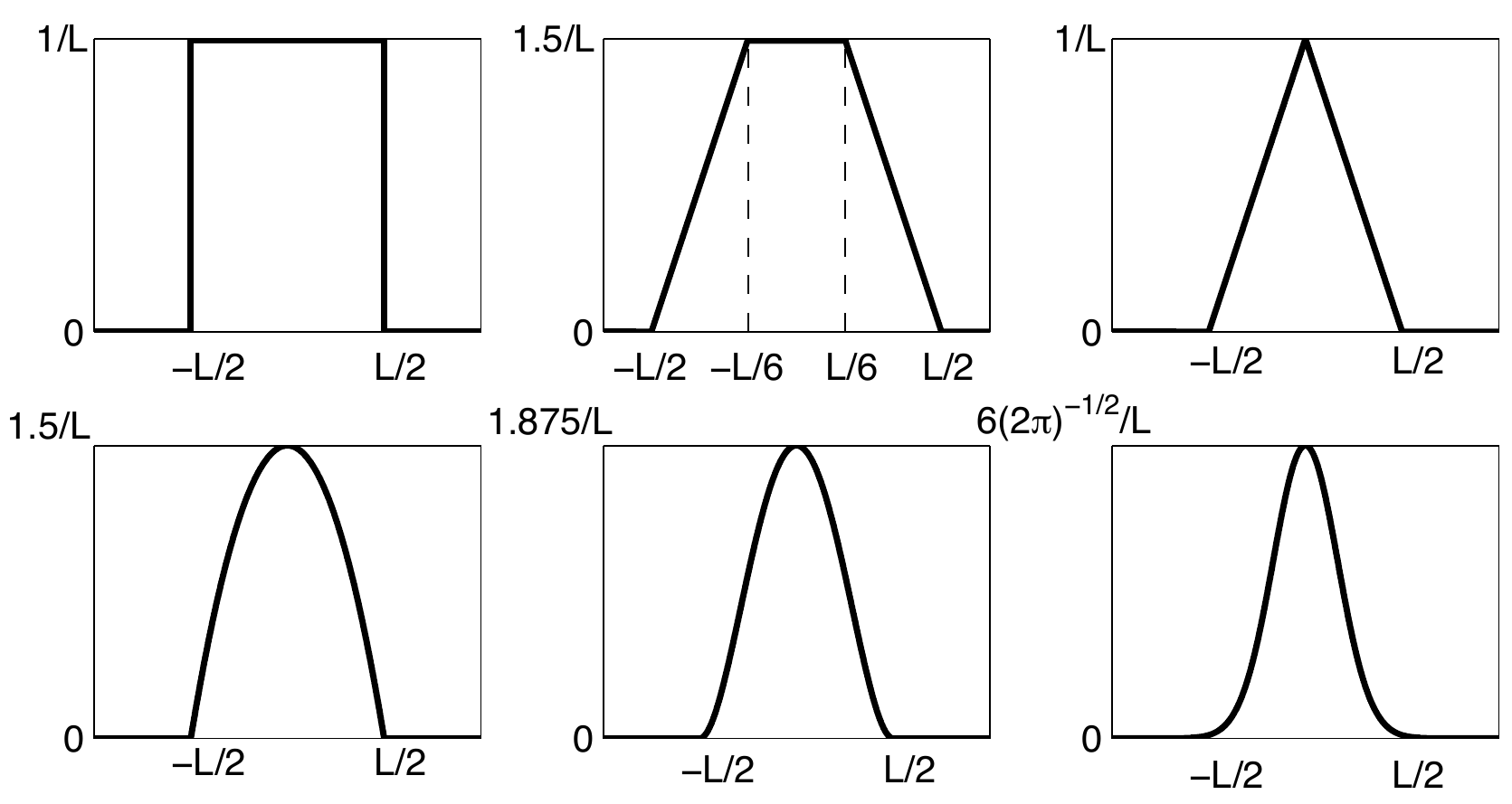}
} \caption{Window functions. Top left: piecewise constant (characteristic) $\psi^{(1)}(x)$; top middle: piecewise linear trapezoidal shape $\psi^{(2)}(x)$;  top right: triangular function $\psi^{(3)}(x)$; bottom left: truncated quadratic $\psi^{(4)}(x)$; bottom middle: truncated $4$th order polynomial  $\psi^{(5)}(x)$;  bottom right: Gaussian $\psi^{(6)}(x)$}
\label{W1}
\end{figure}

\begin{figure}[h] \centerline{
\includegraphics[height=1.3in,angle=0]{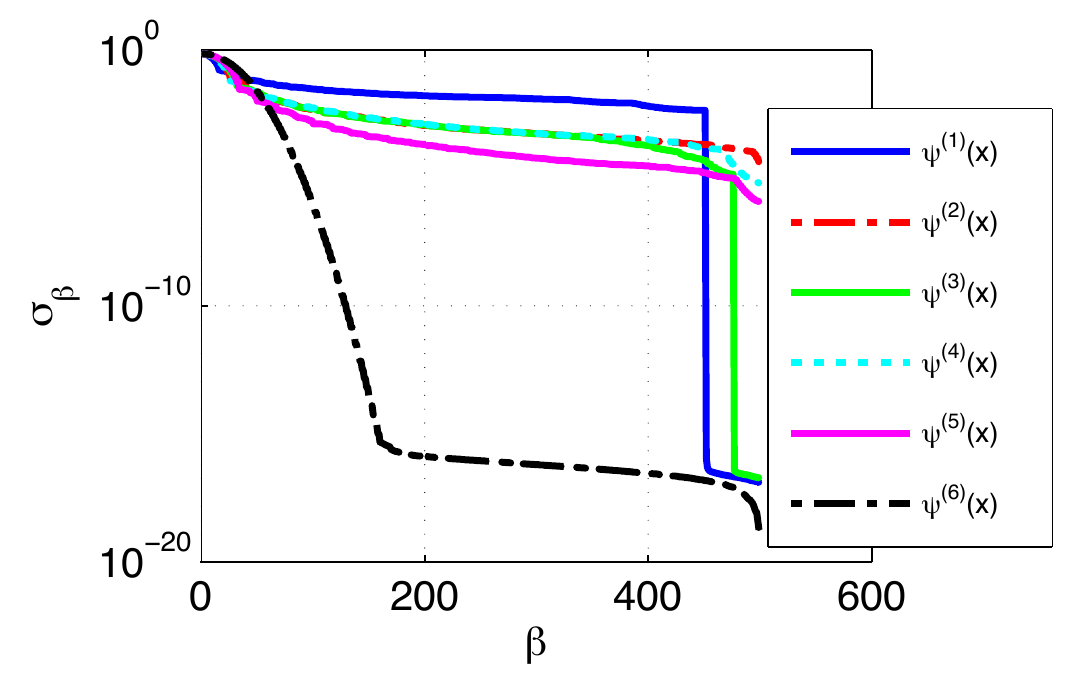}
\hspace{20pt}
\includegraphics[height=1.3in,angle=0]{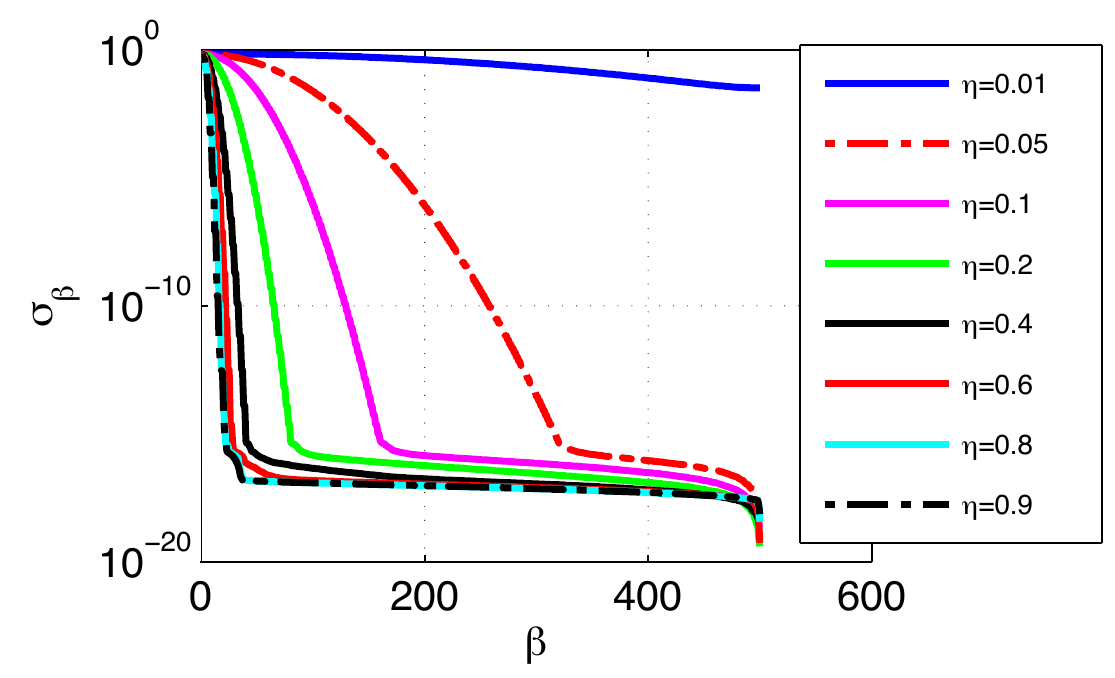}
} \caption{Left panel: singular values for  window functions $\psi^{(i)}$, $i=1,\ldots,6$, with $\eta=0.1$, $N=1000$ and $B=500$. 
Right panel: singular values for  for various $\eta$,  $0.01\leq \eta\leq 0.9$, $\psi^{(6)}$, $N=1000$, $B=500$. In both cases, the graphs of singular values with $N=10000$ and $B=500$ are similar.
}
\label{W2}
\end{figure}

We restrict our attention to the window functions $\psi$ satisfying the conditions
\begin{eqnarray} 
&&\mbox{$\psi$ is non-negative, continuous, and differentiable almost everywhere on the interior of its support and}\label{cond1} \\
&&\int_{-\infty}^\infty\psi(x)=1; \label{cond2}\\[5pt]
&&\psi(x)\to 0 \quad  \mbox{as} \quad |x|\to\infty. \label{cond3}
\end{eqnarray}
%
The function $\psi(x)$ can be compactly supported or fast decreasing as Gaussian. 
We consider several window functions: piecewise constant (characteristic) $\psi^{(1)}(x)$, piecewise linear trapezoidal shape $\psi^{(2)}(x)$, piecewise linear triangular $\psi^{(3)}(x)$, truncated quadratic $\psi^{(4)}(x)$, truncated $4$th order polynomial $\psi^{(5)}(x)$ and Gaussian $\psi^{(6)}(x)$. 
The functions are chosen to have the compact support on $[-L/2,L/2]$. For the  Gaussian function $\psi^{(6)}(x)=\frac{1}{\sqrt{2\pi\sigma}}\exp\left(-\frac{x^2}{2\sigma^2}\right)$, the standard deviation is $\sigma=L/6$, i.e. only about $0.2\%$ of the area under $\psi^{(6)}$ is outside  $[-L/2,L/2]$. Window functions are depicted in Fig. \ref{W1}.
The shifted and rescaled window function $\psi_\eta(\bx-\bq_i(t))$  defines the average properties of the particles in the vicinity of $\bx$. In the case of the  characteristic function $\psi^{(1)}$, the contributions of all particles within a fixed distance of $\bx$ are weighted equally. A better choice of a window function is obtained when the weight decreases to zero as the distance between the particle and the observation point increases. This adds another ``desired" characteristics of a window function (see \cite{Root_Hardy_Swanson_2003}): 
\begin{eqnarray}
&& \psi(x) \quad \mbox{has a maximum at} \quad x=0 \label{cond4}
\end{eqnarray}
Clearly, functions $\psi^{(1)}(x)$ and $\psi^{(2)}(x)$ do not satisfy this property, whereas  the rest of the functions does. 
%
A window function can have a different degree of differentiability, ranging from piecewise constant (as characteristic function) to infinitely many times differentiable as Gaussian. The order of differentiability affects the speed of decay of singular values to zero. In the left panel of Fig.  \ref{W2}  we plot singular values of $\psi^{(i)}(x)$, $i=1,\ldots,6$, for $\eta=0.1$, $N=1000$ and $B=500$.  Results are similar for $N=10,000$. Since we have to invert an operator with kernel $\psi_\eta$, increasing smoothness of the kernel increases ill-posedness of the inverse problem. On a discrete level, the situation is somewhat different. First, discretization itself regularizes the inverse problem (see \cite{Kirsch_1996}). In addition, truncating SVD 
provides additional regularization (see, for example, \cite{Hansen_1987}). 
It is natural to ask how smoothness of the window function effects the reconstruction quality. 
The least smooth function is a characteristic function $\psi^{(1)}(x)$, followed by piecewise linear  trapezoidal $\psi^{(2)}(x)$ 
and triangular function $\psi^{(3)}(x)$. Then we have a truncated quadratic  $\psi^{(4)}(x)$, 
truncated  $4$th order polynomial $\psi^{(5)}(x)$ 
and Gaussian $\psi^{(6)}(x)$.
The singular values of the corresponding matrices are given in the left panel of Fig. \ref{W2}.
As expected, the decay of spectral coefficients is  fastest for the Gaussian and  slowest  for the characteristic function. A sharp drop of singular values for the characteristic function $\psi^{(1)}(x)$ and triangular function $\psi^{(3)}(x)$ indicates that the discrete problem is numerically rank-deficient. The same can be said about the Gaussian that had only about a third of singular values above the machine zero.

If the convolution kernel  $\psi_\eta$ and function $f$ is periodic or extended periodically, then $R_\eta$ is a circular convolution operator (see \cite{Mallat_2009}). If both $f$ and $R_\eta[f]$ are discretized on the same grid, then  the eigenvectors of the circular convolution operator are the discrete complex exponentials and the eigenvalues are Fourier modes of the window function $\psi_\eta$ (see e.g. \cite{Mallat_2009}). Therefore, eigenvectors corresponding to the smallest eigenvalues of the $R_\eta$ carry information about high frequency component of a solution. If $R_\eta[f]$ and $f$ are sampled at different scales, we have to deal with singular value decomposition instead of eigenvalue decomposition. In this case, singular vectors corresponding to the largest singular values would have contribution not only from low frequencies,  but also from some high frequencies. Nevertheless, singular vectors corresponding to smallest singular values will be the most oscillatory and we can still think that singular vectors corresponding to the smallest singular values represent the oscillatory part of a solution. In this regard, the fact that singular values for all window functions but Gaussian decay slowly indicates that solutions corresponding to these window functions will have high frequency component present. Hence,  if there is a numerical error in computation of singular values and singular vectors, especially in those corresponding to the smallest singular values, there may be a significant error in computation of a solution. On the other hand, singular values of a scaled Gaussian window function decay very fast. This means that a fewer singular values can be used to represent a solution. Thus, using a Gaussian is more efficient for large systems of ODEs if the associated error is comparable with other choices of window functions.
%

In order to approximate the stress, one needs to approximate the exact microscopic positions and velocities. 
These approximations are obtained by generating deconvolution approximations of
$J$ and $\tilde v J$, recovering microscopic velocities $\tilde v$ by dividing  $\tilde v J$ by $J$, and reconstructing microscopic positions $\tilde q$ from $J$ using (\ref{Jac}). These are then used to compute approximate stresses. Each of the above steps carries some error, and smaller deconvolution error does not necessarily yield to smaller overall error in approximating stresses $\bT^\eta_{(int)}$ and $\bT^\eta_{(c)}$.

\begin{figure}[h] \centerline{
\includegraphics[height=1.5in,angle=0]{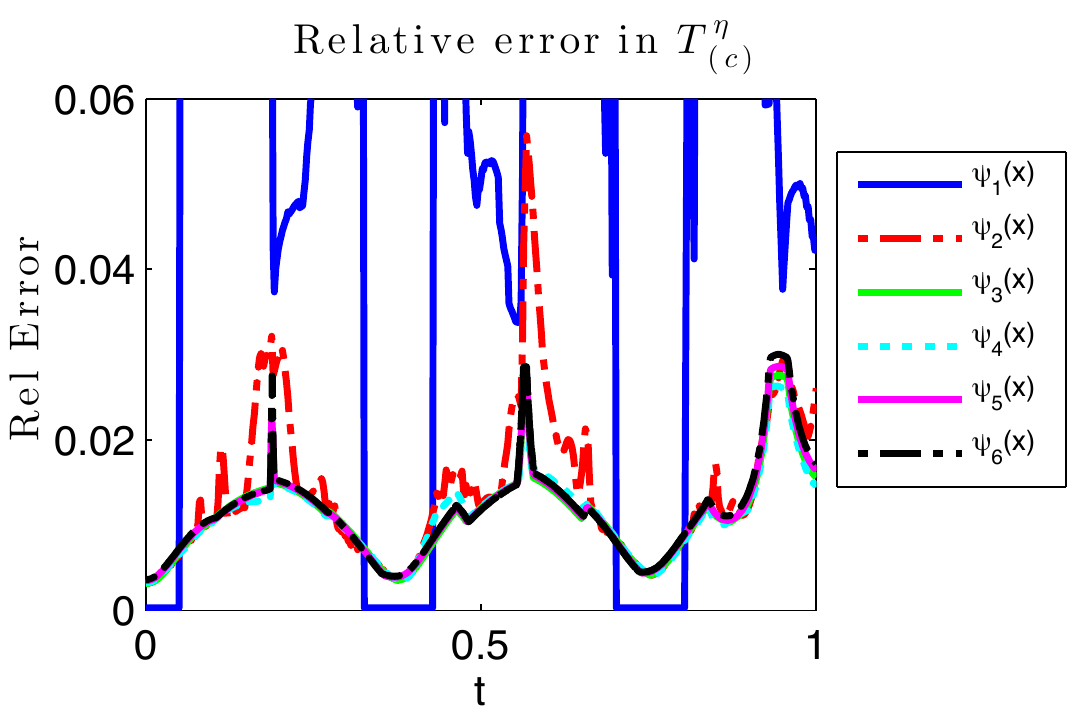}
\hspace{20pt}
\includegraphics[height=1.5in,angle=0]{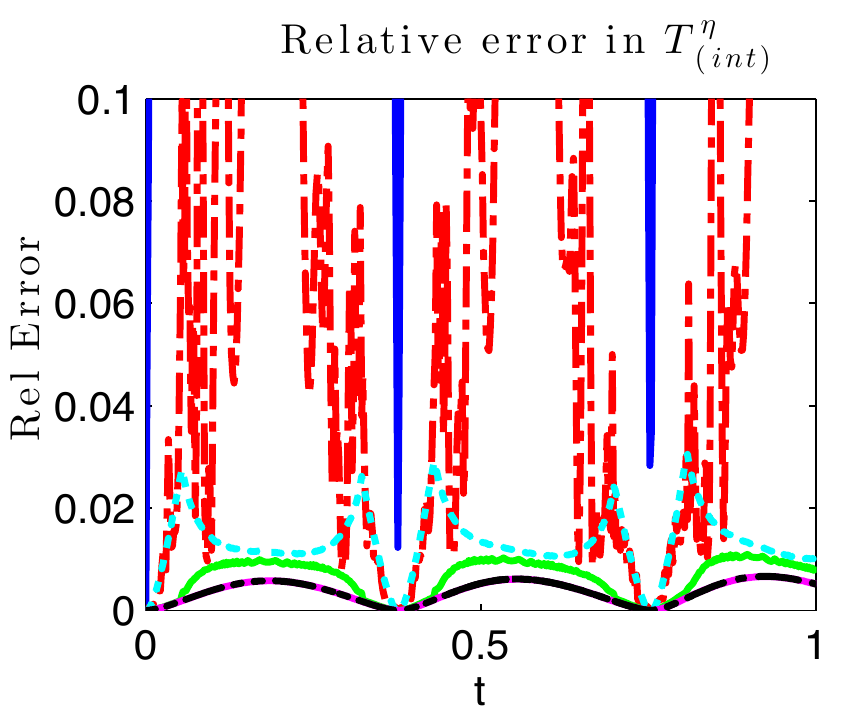}
} 
\caption{Effect of the choice of the window function on the error in approximation of stresses in the first test case with $N=1000$, $B=500$ and $\eta=0.1$. Left panel: convective stress. Right panel: interaction stress.
}
\label{Fsine_var_fm_T1}
\end{figure}

In Fig. \ref{Fsine_var_fm_T1} we compare the $l^\infty$-relative errors in approximation of the convective $T^\eta_{(c)}$ and interaction $T^\eta_{(int)}$ stresses for the first test case. 
Clearly, the characteristic function $\psi^{(1)}$ has the worst performance. The absolute error (not shown here) in $T^\eta_{(c)}$ is at most $10^{-7}$ or $13\%$, while the error in using other window functions is much smaller. For example, piecewise linear $\psi^{(2)}$ gives the absolute error of two order lower: at most $2.5\cdot 10^{-9}$ or $6\%$, while the $4$th order truncated polynomial $\psi^{(5)}$ and Gaussian $\psi^{(6)}$ give the  error $10^{-9}$ or only $2-3\%$. The error in using the rest of the window functions is between $10^{-9}$ and $2.5\cdot 10^{-9}$ or between $3$ and $6\%$. The right panel of Fig. \ref{Fsine_var_fm_T1} suggests that the approximation of  $T^\eta_{(int)}$ with $\psi^{(1)}$ has too large error, which is not acceptable.  The error with $\psi^{(2)}$ is at most $4\cdot 10^{-2}$ (reaching $75-100\%$ at some times) and drops to $2\cdot 10^{-4}$ (or $0.5\%$) with $\psi^{(5)}$ and $\psi^{(6)}$. Functions $\psi^{(3)}$ and $\psi^{(4)}$ give $1-3\%$ error.

The worse behavior of $\psi^{(1)}$, followed by  $\psi^{(2)}$, could be explained by the fact that function $\psi^{(1)}$ is not even continuos everywhere that violates condition (\ref{cond1}), while $\psi^{(2)}$ is only piecewise continuous. Moreover, both functions  do not have a strict maximum at $x=0$, thus violating the condition (\ref{cond4}).  For the rest of the window functions considered here, all conditions (\ref{cond1})--(\ref{cond4}) are satisfied.

\begin{figure}[h] \centerline{
\includegraphics[height=1.5in,angle=0]{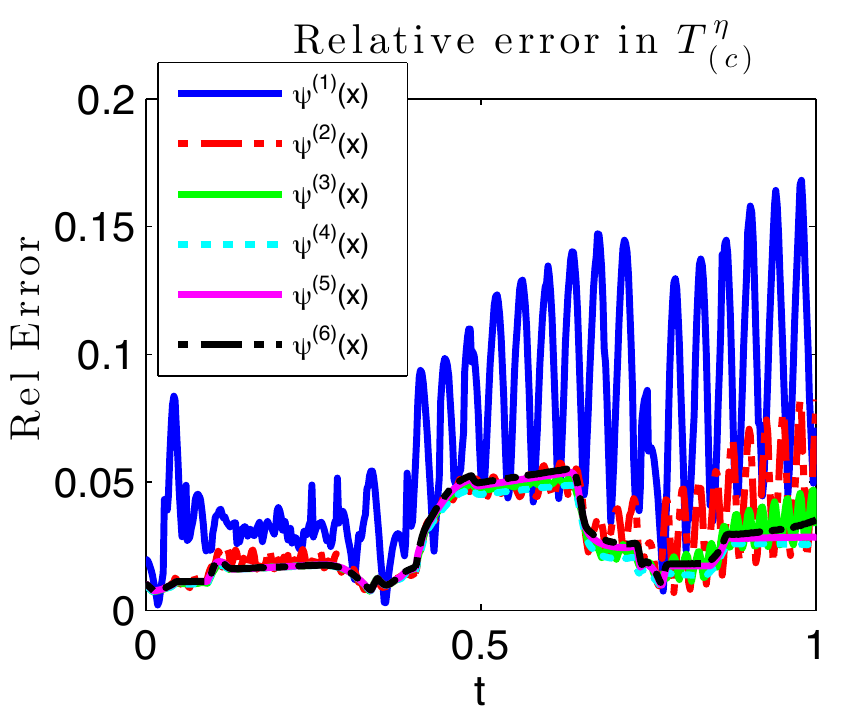}
\hspace{20pt}
\includegraphics[height=1.5in,angle=0]{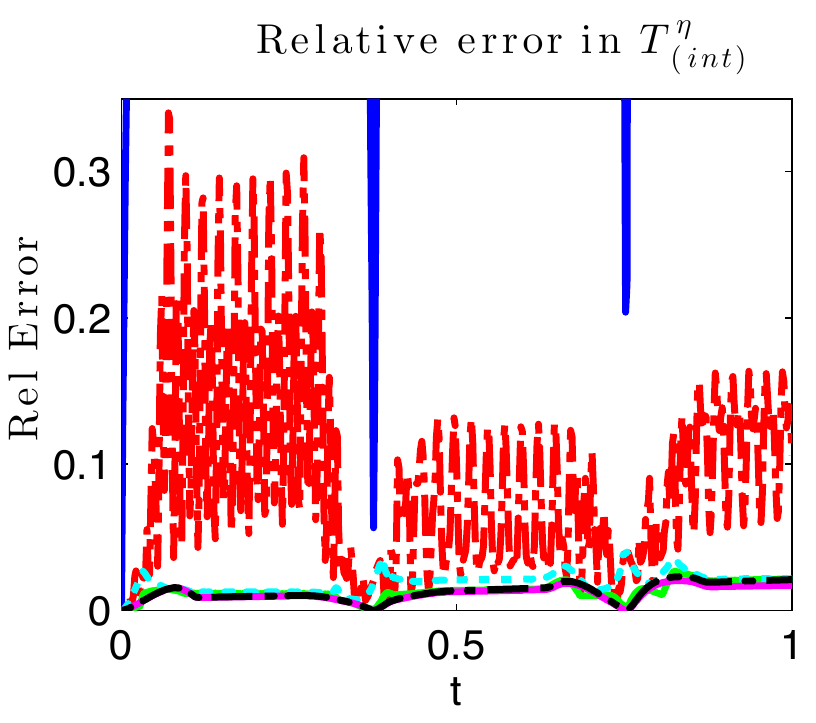}
} \caption{The effect of the choice of the window function on the error in approximation of stresses in the second test case. The relative error is shown for the case with $N=1000$, $B=500$ and $\eta=0.1$. Left panel: convective stress. Right panel: interaction stress.
}
\label{W1b}
\end{figure}

The results for the second test case are shown in Fig.  \ref{W1b}. Again, the worst performance is by $\psi^{(1)}$, followed by $\psi^{(2)}$. The best results are obtained with $\psi^{(5)}$ and $\psi^{(6)}$, the rest of the functions give intermediate results. Specifically, the absolute error in approximating the convective stress  $T^\eta_{(c)}$ using $\psi^{(1)}$ is at most $5\cdot 10^{-6}$ or $17\%$, it drops to $1.5\cdot 10^{-6}$ or $8\%$ with $\psi^{(2)}$ and reaches only $0.5-0.75 \cdot 10^{-6}$ or $2-5\%$ with $\psi^{(3)}-\psi^{(6)}$. We also note that the error with functions $\psi^{(1)}-\psi^{(3)}$ is highly oscillatory while it is not oscillatory with the rest of the functions.  The absolute error in approximation of the interaction stress $T^\eta_{(int)}$ using function $\psi^{(2)}$ is at most $10^{-2}$ or $30\%$ (decreases to $15\%$ at later times). Function $\psi^{(4)}$ gives $10^{-3}$ absolute error or $4\%$, while functions $\psi^{(3)}$, $\psi^{(5)}$ and $\psi^{(6)}$ produce only $0.6-0.7\cdot 10^{-3}$ absolute error or at most $1-2\%$.

The experiments in this section show that the most accurate approximation of the stresses is obtained using either truncated $4$th order polynomial function $\psi^{(5)}$ or Gaussian $\psi^{(6)}$. The absolute error is typically smaller with $\psi^{(6)}$ but  the relative error is sometimes slightly smaller with $\psi^{(5)}$. However,  the difference in performance of these functions is not significant. At the same time with $\psi^{(5)}$ and, for example, with $N=10,000$ and $B=500$, one has to use all $500$ singular values and singular vectors since all singular values are above the threshold $\sigma_\alpha=10^{-13}$ (they decay to at most $10^{-6}$, see the left panel  of Fig. \ref{W2}), and only $147$ singular values with $\psi^{(6)}$, which is more efficient for large systems of particles. For this reason, in what follows we fix Gaussian window function $\psi^{(6)}$ and study the effect of other parameters such as averaging width $\eta$ and scale separation.

\section{Choice of mesoscale resolution parameter $\eta$} \label{resolution_parameter}

The parameter $\eta$  in  (\ref{density}), (\ref{mom}) determines the size of the averaging region and the
amount of high frequency filtering. 
For smaller $\eta$, there is less damping of high frequency content of the solution, while using a larger $\eta$ produces smoother and smaller averages. The latter is discussed in Section \ref{spectral_evolution}. 
As $\eta$ increases, the singular values the corresponding matrix $A^\eta$ decay at a higher rate. This can be seen in the right panel of Fig. \ref{W2} where we show  the singular values for $N=1000$ and $B=500$ and $\eta=0.01, 0.05, \ldots,0.9$. (The results with $N=10000$ and $B=500$ are similar since they are more sensitive to the choice  of $B$ and not of $N$.)

On the one hand, for larger $\eta$, the average contains less high frequency information, and it may be more difficult to reconstruct the same microscopic solution from increasingly smoothed averages. This may increase the error in computation of $Q_\eta$.
On the other hand, 
computation of the stress involves another averaging (the outer layer $R_\eta$ in (\ref{sandwich})) that may decrease the overall error even if $J$ and $\tilde\bv$ are recovered more poorly.
In this section, we study the cumulative effect of these two competing tendencies on the overall relative error in approximating stress.

\begin{figure}[h] \centerline{
\includegraphics[height=1.5in,angle=0]{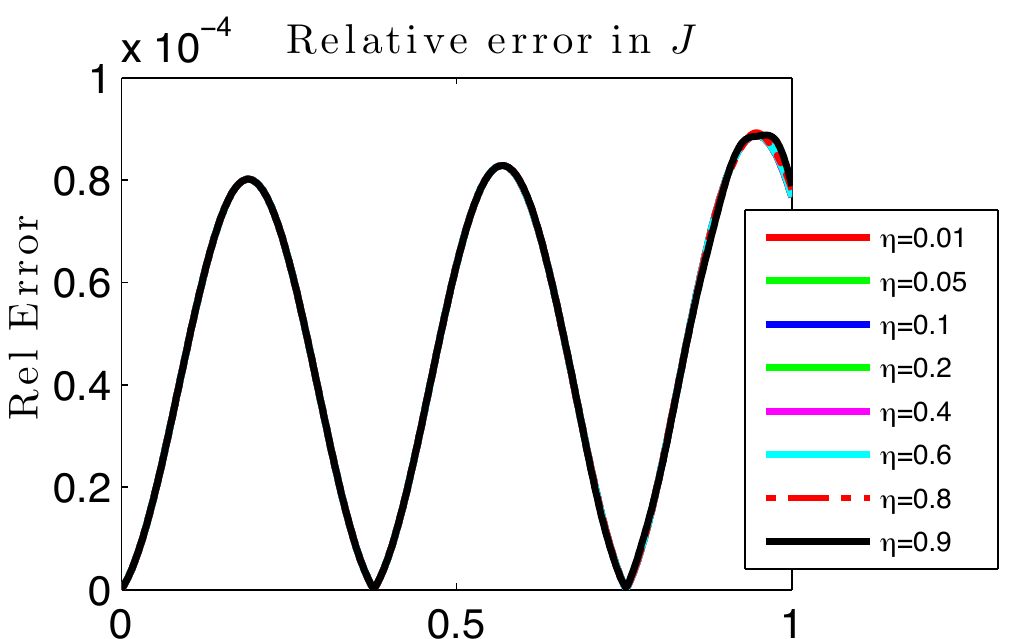}
\hspace{20pt}
\includegraphics[height=1.5in,angle=0]{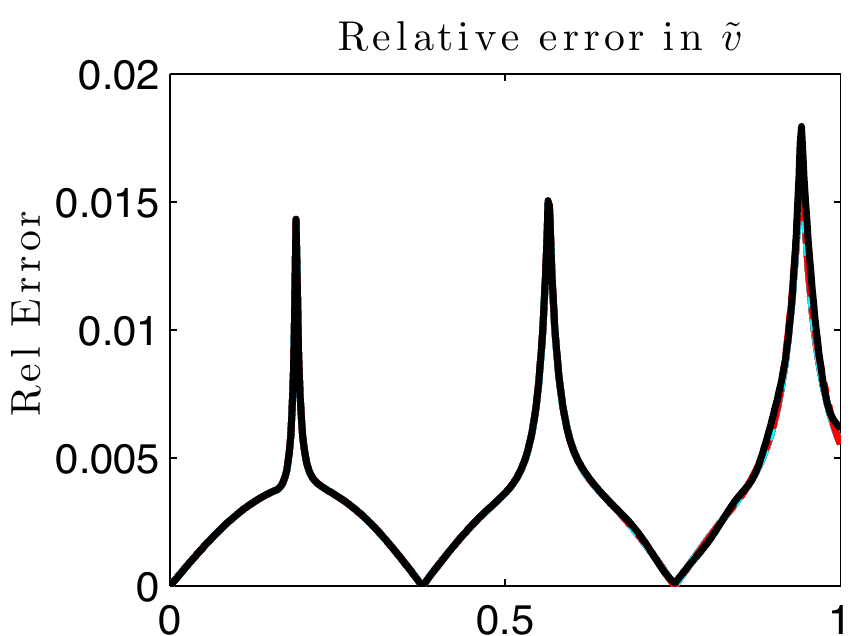}
} \caption{The effect of the choice of the resolution parameter $\eta$ on the error in approximation of the Jacobian (left panel) and velocity (right panel) for the first test case with $N=1000$ and $B=500$.
}
\label{W8a1}
\end{figure}

To see how the choice of $\eta$ affects the reconstruction of the Jacobian and microscopic velocity, and subsequent stress approximation, we fix $N=1000$, $B=500$ and the window function $\psi^{(6)}(x)$, and
vary $\eta$ between $10^{-2}$ and $0.9$.

\subsection
{First test case}
\begin{figure}[h] \centerline{
\includegraphics[width=2.0in,angle=0]{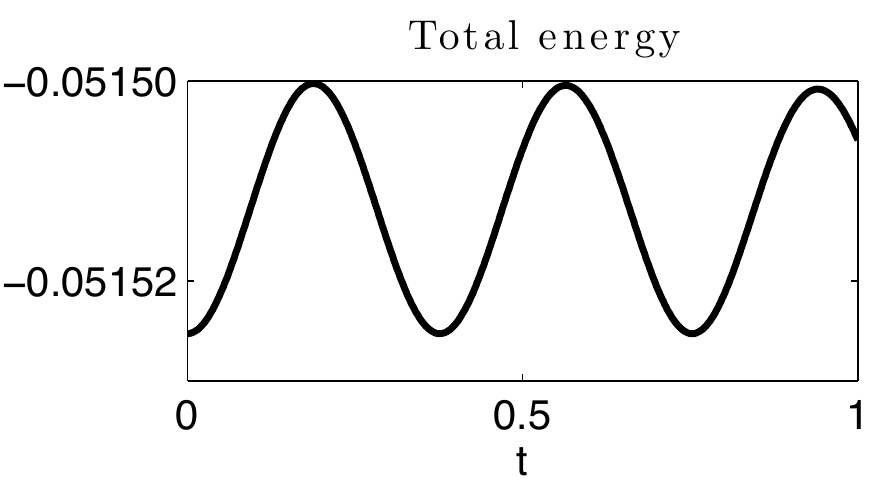}
\hspace{20pt}
\includegraphics[width=2.0in,angle=0]{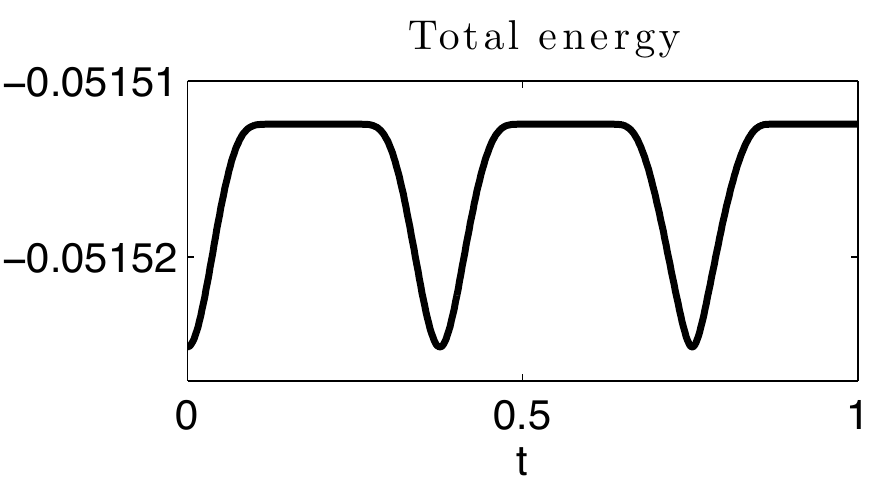}
} \caption{Evolution of the total energy for $N=1000$,  $B=500$ and $\eta=0.1$. Left panel:  the first test case. Right panel:  the second test case.}
\label{W8b}
\end{figure}

In this case, the errors in the Jacobian and velocity approximation, shown in  Fig. \ref{W8a1}, are essentially independent of $\eta$, though there is a slight dependence on $\eta$ for times after $t=0.8$.   
Both absolute and relative errors in the Jacobian oscillate in a quasiperiodic manner. The amplitude of oscillations in absolute error increases slightly with time and reaches $9\cdot 10^{-5}$ at most during the simulation time, which is less than $0.01\%$ error.   The behavior of the error can be connected to the evolution of  the total computed energy of the system shown in left panel of Fig. \ref{W8b}. As can be seen from the graph, the energy is not completely conserved. Instead it
oscillates periodically and deviates from its value at $t=0$ (the true energy of the system) by at most $2.5057\cdot 10^{-5}$ (less than $0.05\%$ change). 
In the regions where the energy starts deviating from its initial value, the error in Jacobian reconstruction increases. When the energy comes back to its initial value, the error in Jacobian also decreases.  The relative error in the Jacobian reconstruction is similar to the absolute error since the exact Jacobian has values close to $1$. The error in velocity  reconstruction,  shown in the right panel of Fig. \ref{W8a1}, has a more nonlinear dynamics. Similarly to the error in the Jacobian, the error in velocity increases at those times when the total energy starts deviating more from its initial value and decreases when the total energy comes back to this value. The maximum absolute error during the simulation time is under $3.5\cdot 10^{-5}$ which is $1.8\%$ error.

\begin{figure}[h] \centerline{
\includegraphics[height=1.5in,angle=0]{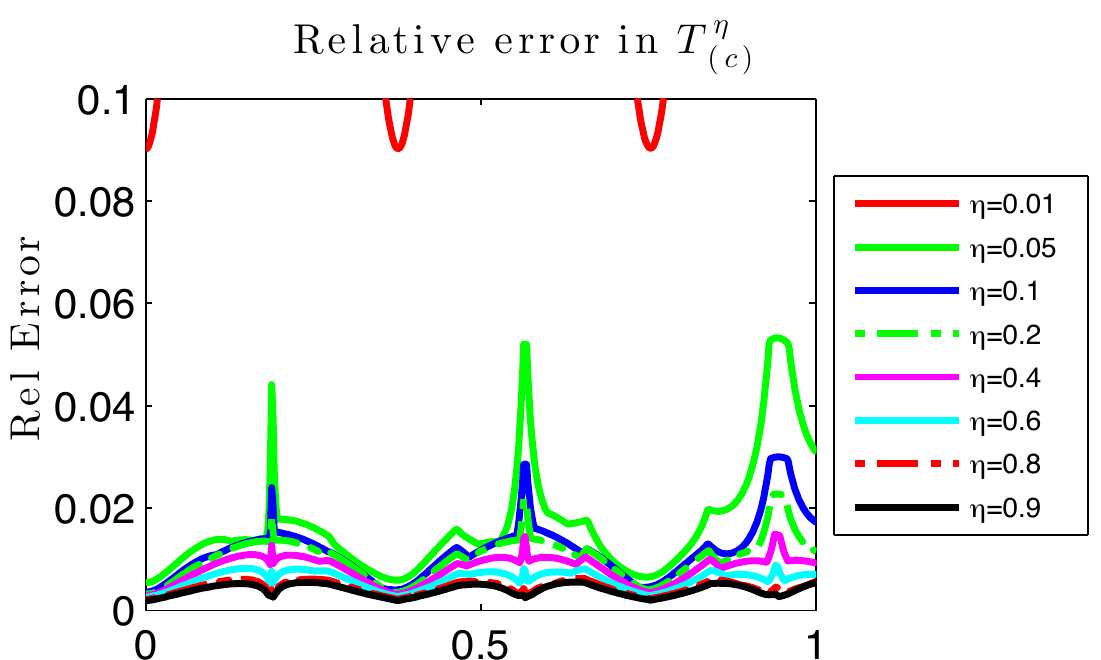}
\hspace{20pt}
\includegraphics[height=1.5in,angle=0]{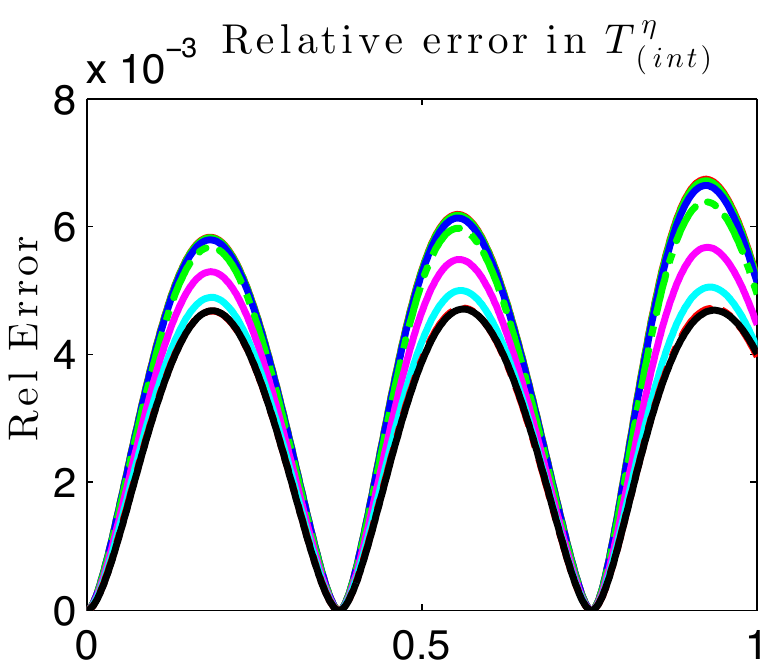}
} \caption{
The effect of the choice of the resolution parameter $\eta$ on the error in approximation of stresses in the first test case with $N=1000$ and $B=500$. Left panel: convective stress. Right panel: interaction stress.
}
\label{W8a3}
\end{figure}

The errors in both convective and interaction stresses  depend on $\eta$ as can be seen from Fig. \ref{W8a3}. 
While the absolute error in $T^\eta_{(c)}$ is not monotonic in $\eta$ at all times, it is typically larger for larger $\eta$.  The time oscillations of the error do not exceed $1.5\cdot 10^{-7}$ . It is the largest with $\eta=0.8$, followed by $\eta=0.6$ and $\eta=0.9$. However, the relative error, shown in the left panel of Fig. \ref{W8a3}, decreases monotonically as $\eta$ increases. More specifically, the error with $\eta=0.01$ is the largest and varies between $9$ and $100\%$, whereas  it is the smallest with $\eta=0.9$ and it varies between  $0.3$ and $0.6\%$. For example, for  $\eta=0.1$, the error fluctuates between $0.5$ and $3\%$.
Both absolute and relative errors in  approximation of $T^\eta_{(int)}$  oscillate in time and depend monotonically on $\eta$: they decrease as $\eta$ increases. The latter is shown in the right panel of Fig. \ref{W8a3}. The absolute error is of the order of $10^{-4}$ and reaches $0.7\%$ at most for $\eta=10^{-2}$ and less than $0.5\%$ for the largest $\eta=0.9$.

\subsection
{Second test case} 
\begin{figure}[h] \centerline{
\includegraphics[height=1.5in,angle=0]{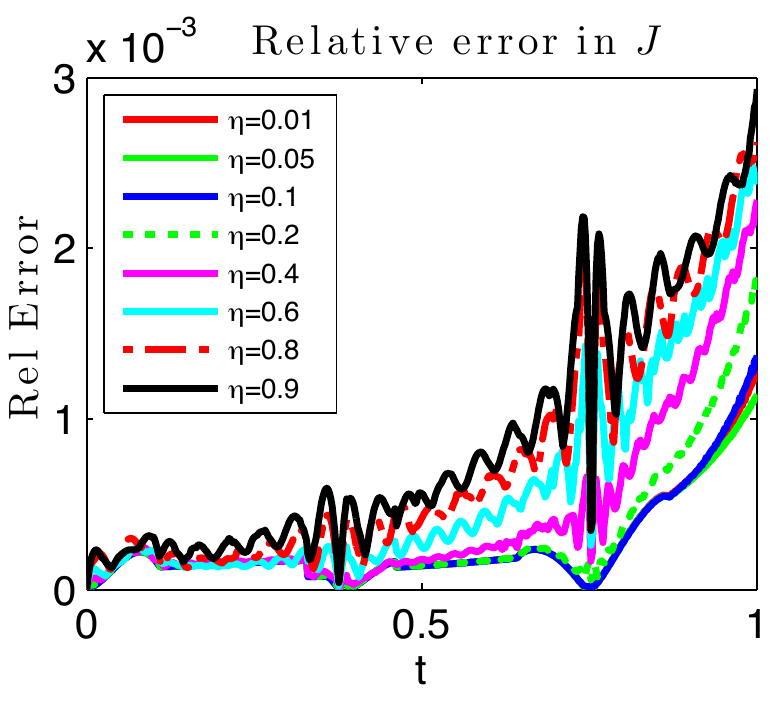}
\hspace{20pt}
\includegraphics[height=1.5in,angle=0]{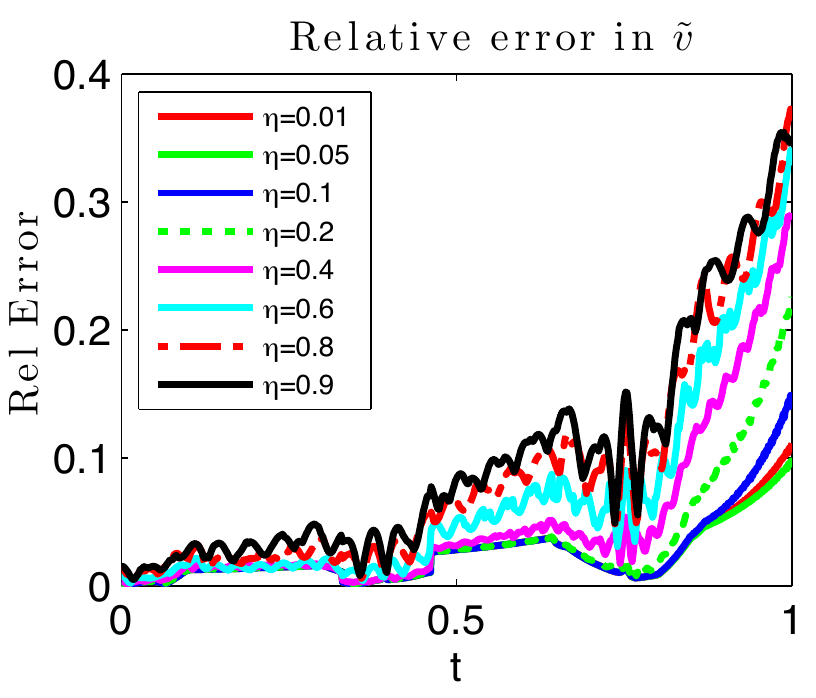}
} \caption{Effect of the choice of $\eta$ on the error in reconstruction of Jacobian (left panel) and velocity (right panel) in the second test case with $N=1000$, $B=500$, $\eta=0.01, \ 0.05, \ 0.1,\ldots, 0.9$.
}
\label{W8a5}
\end{figure}

\begin{figure}[h] \centerline{
\includegraphics[height=1.5in,angle=0]{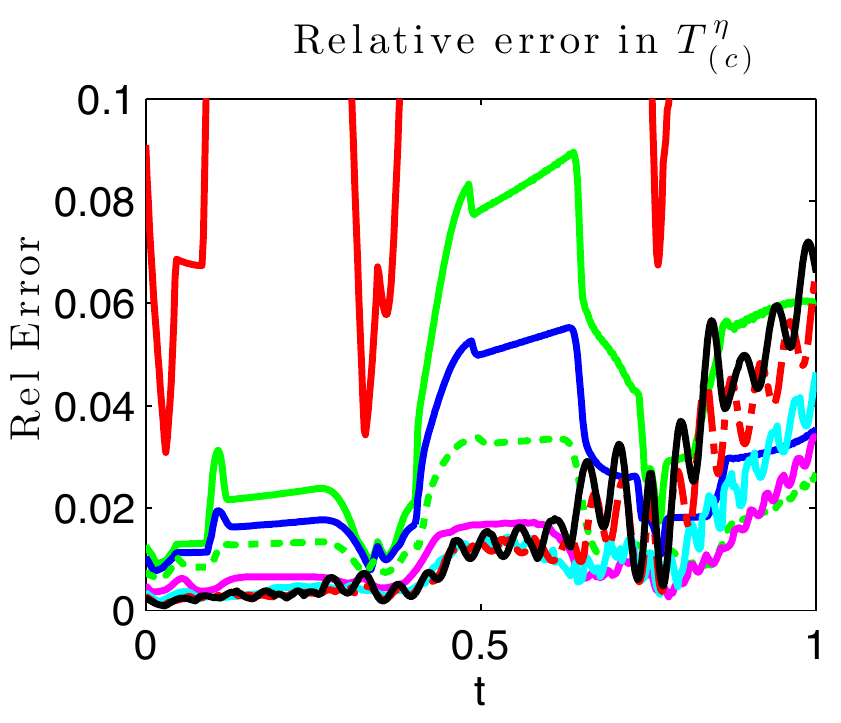}
\hspace{20pt}
\includegraphics[height=1.5in,angle=0]{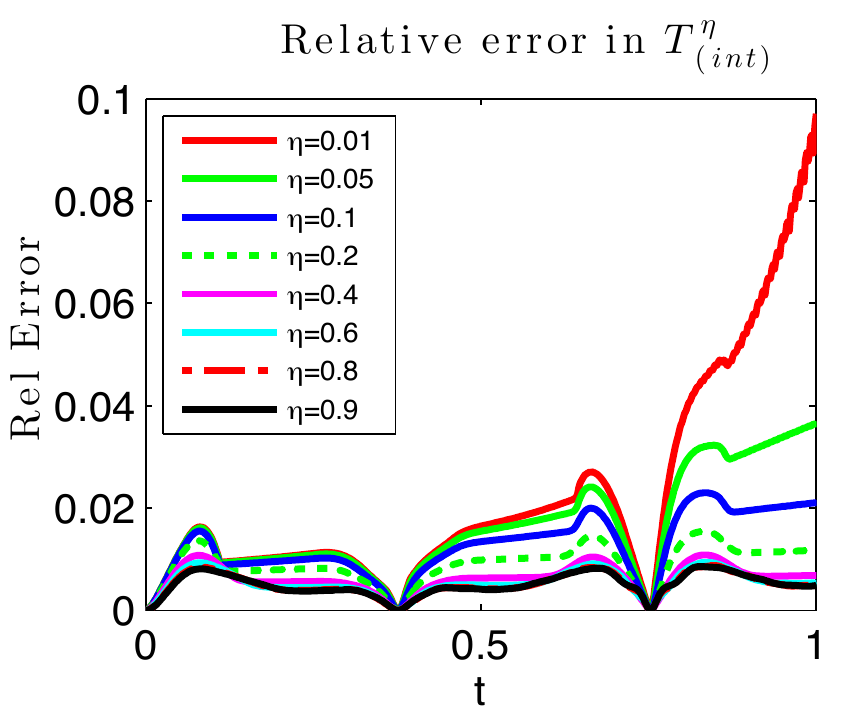}
} \caption{Effect of the choice of $\eta$ on the error in stress approximation in the second test case with $N=1000$, $B=500$, $\eta=0.01, \ 0.05, \ 0.1,\ldots, 0.9$. Left panel: convective stress. Right panel: interaction stress.}
\label{Fa7}
\end{figure}

The simulation results with various $\eta$ are presented in Figs. \ref{W8a5} and \ref{Fa7}. Fig. \ref{W8a5}  indicates that the reconstruction of both Jacobian and velocity gets worse as $\eta$ and time $t$ increase. However, the situation with stress approximation is different. The error in approximating the interaction stress is the smallest with the largest $\eta=0.9$ used. It is below $1\%$ during the entire simulation time, whereas the error with the smallest $\eta=0.01$ reaches $10\%$ by the end of simulations and it is the largest among all values $\eta$ used. As for the convective stress, the error with $\eta=0.9$ is the smallest at early times (around $0.2\%$), then starts increasing and it is of the same order as with $\eta=0.6$ and $0.8$. After $t=0.6$, the errors with $\eta=0.4$ and $0.2 $ are the smallest (up to $3\%$) and errors with $\eta=0.6$, $0.8$ and $0.9$ reach $4$ to $7\%$. It should be noted though that the convective stress is much smaller than the interaction stress (by at least two orders of magnitude) and the absolute errors in approximating the convective stress are at the order of $10^{-6}$ for all $\eta$ used in the experiments.

Consider, for example, $\eta=0.1$. From Fig. \ref{W8a5} we can see that the error in Jacobian and velocity approximation is $0.02\%$ and $2\%$ at early times and reaches $0.12\%$ and $15\%$ by the end of simulations, respectively. At the same time, the error in the convective stress varies between $1-2\%$ and $5\%$ during the simulation period. The error in the interaction stress does not exceed $2\%$.  As we can see, the error in the approximation of the Jacobian stays small during the simulation period, though it grows slowly with time (at most linearly). Despite of the large error in the approximation of velocity, we still get very good stress approximation.  The situation is similar for other values of $\eta$.

%

\section{Spectral evolution of averages and stresses} \label{spectral_evolution}

As was mentioned in Section \ref{resolution_parameter},  parameter $\eta$ determines the size of the averaging window and the amount of high frequency filtering. With larger $\eta$, the averages are smoother and smaller. 
This can be seen by considering the Fourier transform of an average 
as follows. 

Recall a typical one-particle dynamical function used in statistical mechanics
\begin{equation}
\label{av1}
{g}_{sm}(t, \bx)=\sum_{i=1}^N g(\bq_i(t),\bv_i(t)) \delta(\bx-\bq_i(t)),
\end{equation}
where $\delta$ is delta-distribution. The Fourier transform of ${g}_{sm}$ with respect to $\bx$ is
\begin{equation}
\label{av2}
\widehat{{g}}_{sm}(t, \bxi)=\sum_{i=1}^N g(\bq_i(t),\bv_i(t)) e^{i\bxi\cdot \bq_i}.
\end{equation}
Now compare this with a windowed spatial average
\begin{equation}
\label{av3}
\overline{g}(t, \bx)=\sum_{i=1}^N g(\bq_i(t),\bv_i(t)) \psi_\eta(\bx-\bq_i(t)),
\end{equation}
and the corresponding Fourier transform
\begin{equation}
\label{av4}
\widehat{\overline{g}}(t, \bx)=\sum_{i=1}^N g(\bq_i(t),\bv_i(t)) \hat{\psi}(\eta\bxi)e^{i\bxi\cdot \bq_i}= \hat{\psi}(\eta\bxi)\widehat{{g}}_{sm},
\end{equation}
where $\hat\psi$ denotes the Fourier transform of $\psi$.  Thus, the Fourier transform of $\overline{g}$ is obtained from $\widehat{g}_{sm}$ by low-pass filtering
(multiplication by $\hat\psi(\eta\bxi)$).
If  $\lim_{|\bk|\to\infty}|\hat\psi(\bk)|=0$, which is true for any $L_1$ function by Riemann -- Lebesgue Lemma, then $\hat\psi(\eta\bxi)$ converges to $0$ for each $\bxi\not=0$ as $\eta\to\infty$.
Since $\widehat{{g}}_{sm}$ does not depend on $\eta$, equation (\ref{av4}) implies $\widehat{\overline{g}}$ approaches $0$.
The rate of decay of $\hat\psi$ increases with smoothness as $\psi$. Thus, increasing $\eta$ produces progressively more filtered versions of $\widehat{g}_{sm}$. 

Flexibility afforded by varying $\eta$ is convenient for studying large scale behavior of the averages. For example, in statistical physics, the derivation of hydrodynamical equations and computation of fluid viscosity by Green-Kubo formulas 
(see e.\,g. \cite{Berne_1977}) employs truncated Taylor expansions of 
$\widehat g_{sm}$ at $\bxi=0$. Estimating the error of these approximations for large $\bxi$ may be difficult, while multiplication by $\widehat\psi(\eta\bxi)$ makes analysis easier. 
Another useful feature of (\ref{av4}) is the possibility to adjust the size of the low-frequency neighborhood of interest by changing $\eta$.

\begin{figure}[h] \centerline{
\includegraphics[width=3.5in,angle=0]{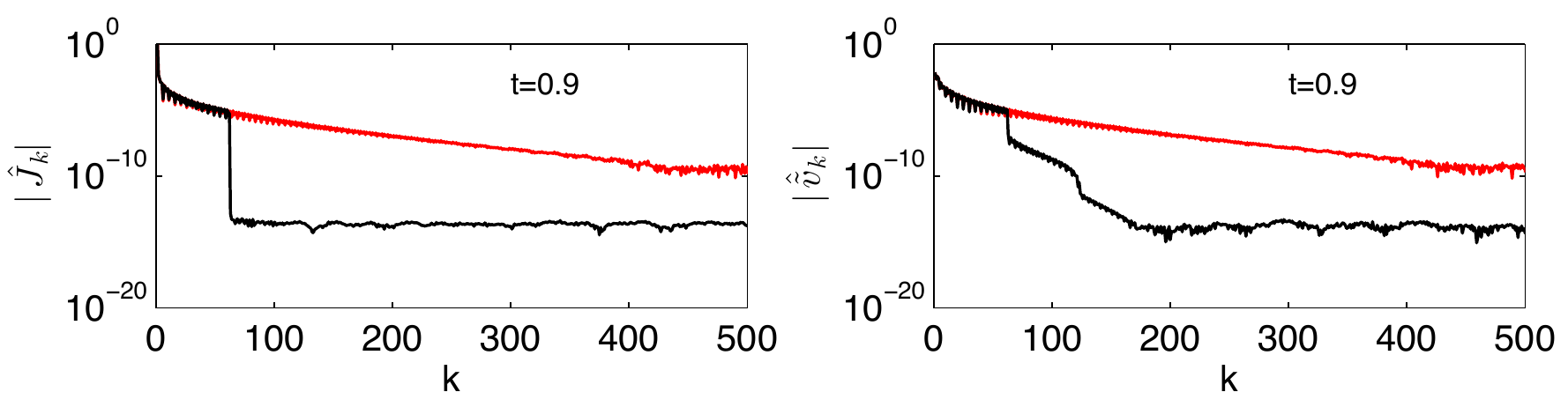}
} \caption{Discrete Fourier coefficients of the Jacobian (left panel) and velocity (right panel) for the second test case with $N=10,000$, $B=500$,  $\eta=0.1$. The logarithm of coefficients' amplitudes is plotted against wavenumber $k$ at $t=0.9$.
%
The red curves are exact solutions, black -- approximations.}
\label{WDFT1}
\end{figure}

\begin{figure}[h] \centerline{
\includegraphics[width=3.5in,angle=0]{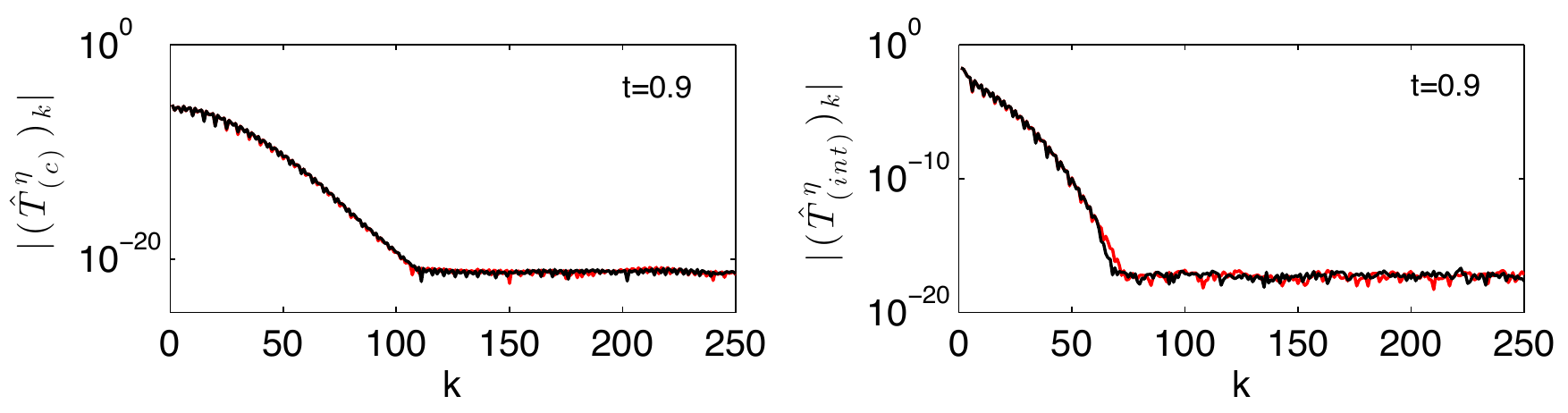}
} \caption{Discrete Fourier coefficients of the convective stress (left panel) and interaction stress (right panel) for the second test case with $N=10,000$, $B=500$, $\eta=0.1$ and $t=0.9$. The red curves are exact solutions, black -- approximations.}
\label{WDFT2}
\end{figure}

To investigate quality of our approximations in the Fourier space,
we analyze Fourier spectra of the exact Jacobian, velocity, stresses  and their approximations. 
We consider only the second test case because the initial velocity in this case has full spectrum unlike the first test case.
%
%
%
Since the approximation tends to become worse at later times, we show the relevant spectra at the ``worst case scenario" time $t=0.9$.  Fig. \ref{WDFT1} depicts the spectra of the exact Jacobian and velocity (red curves) and their approximations (black curves).
The left panel of Fig. \ref{WDFT1} indicates that we only capture about $70$ first low frequency modes of the Jacobian. 
Similarly, the first $70$ modes of the velocity are well reconstructed, while modes between $70$ and $180$ have much smaller amplitudes than in the exact velocity. While spectra of both Jacobian and velocity are not very accurate, 
spectral approximations of both convective and interaction stresses are quite good (see Fig. \ref{WDFT2}). Both exact stresses have only low frequency components  ($110$ for the convective stress and only $70$ for the interaction) and all these modes are  captured perfectly! This demonstrates that it is not necessary to recover higher frequency modes of the Jacobian and velocity in order to approximate stress accurately. 

\begin{figure}[h] \centerline{
\includegraphics[width=3.5in,angle=0]{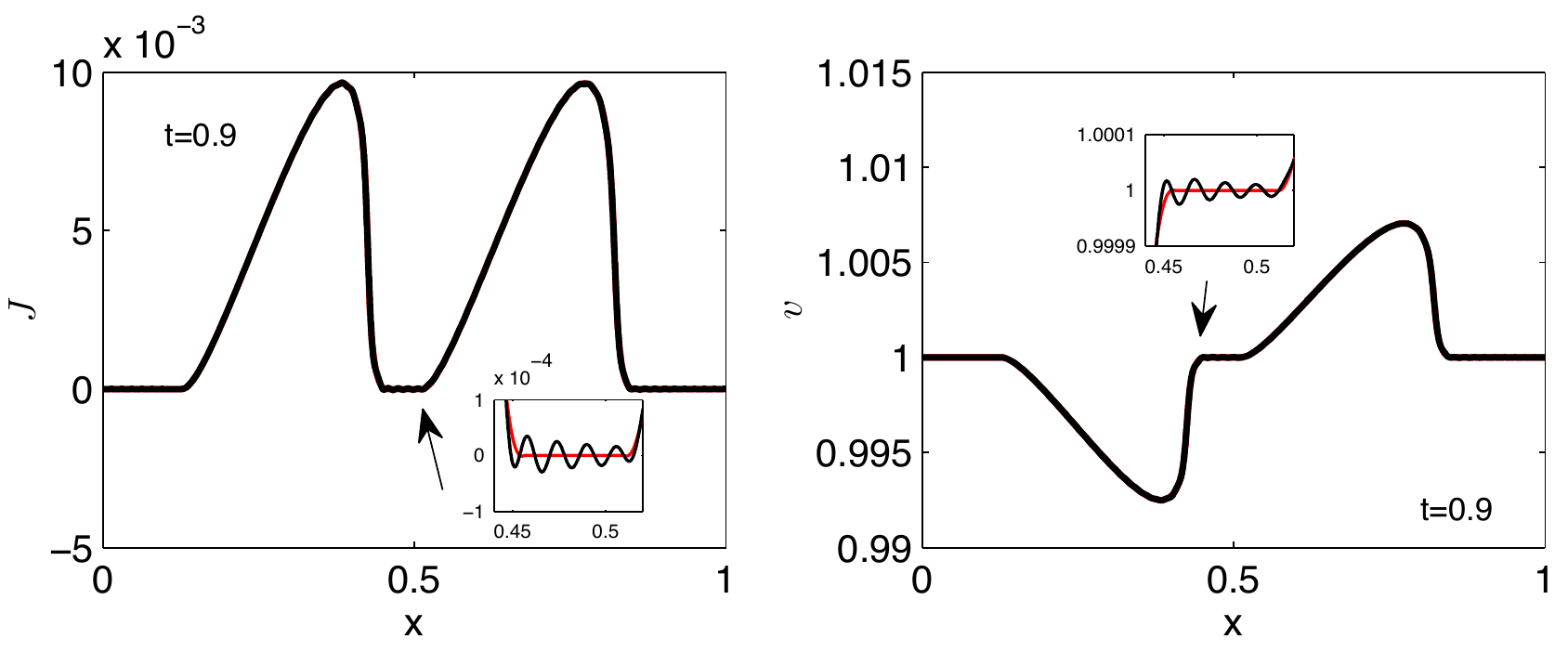}
} \caption{Reconstruction of the Jacobian (left panel) and velocity (right panel) in the second test case with $N=10,000$, $B=500$,  $\eta=0.1$. Exact (red curves) and approximate (black curves) solutions are shown at $t=0.9$.}
\label{WDFT5}
\end{figure}

\begin{figure}[h] \centerline{
\includegraphics[width=3.5in,angle=0]{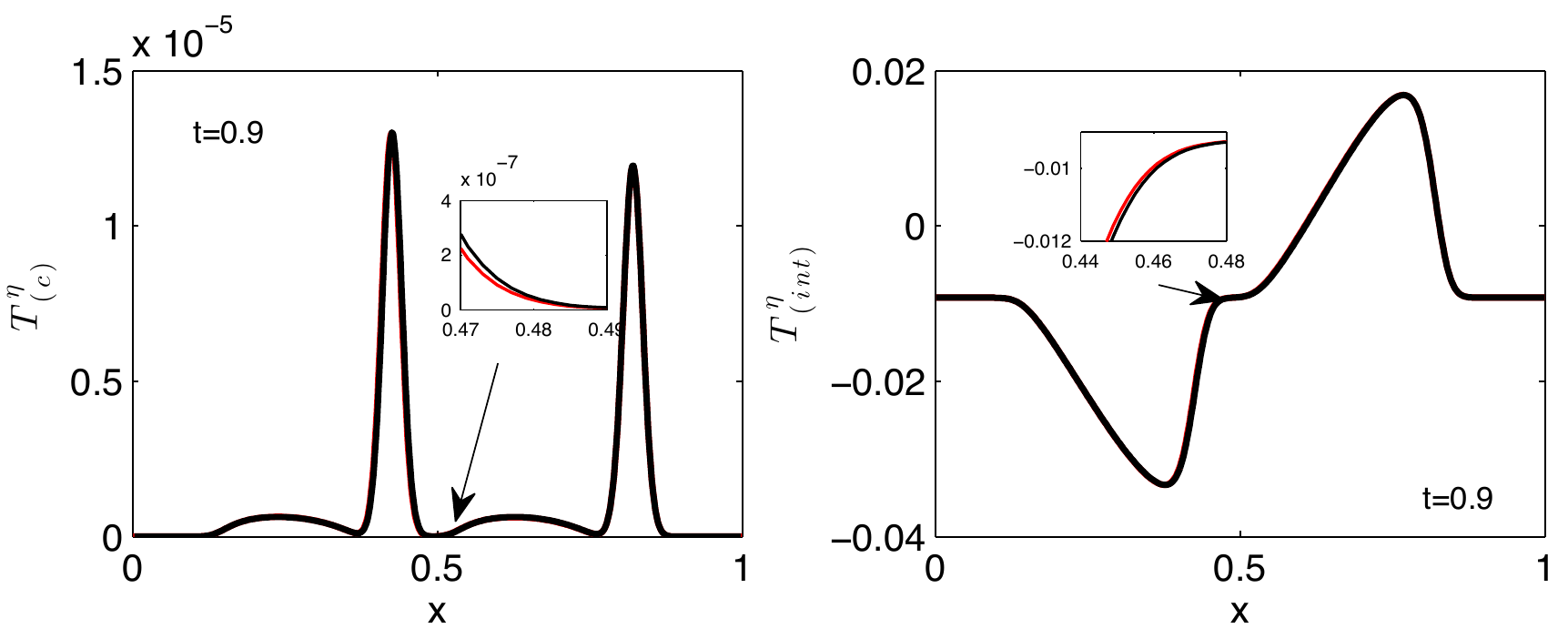}
} \caption{Reconstruction of the convective stress (left panel) and interaction stress (right panel)  in the second test case with $N=10,000$, $B=500$ and $\eta=0.1$. Exact and approximate solutions are shown at $t=0.9$.   The red curves are exact solutions, black -- approximations.}
\label{WDFT6}
\end{figure}

Loss of accuracy in using truncated spectra for deconvolution often leads to Gibbs phenomenon. It is indeed present in both Jacobian and velocity reconstructions shown in 
Fig. \ref{WDFT5}. The amplitude of Gibbs ripples seem to increase with $\eta$. In contrast, the approximations to stresses do not suffer from Gibb's oscillations as can be seen from Fig. \ref{WDFT6}.
Gibb's phenomenon is typical for solutions of linear systems using a truncated SVD approach (see \cite{Boyd_2002, Bruno_2003, Bruno_Han_Pohlman_2007, Boyd_Ong_2009}). In \cite{Lyon_2012}, Gibb's phenomenon is controlled by using Sobolev smoothing. In our case smoothing is done naturally by averaging present in the stress approximation (see (\ref{sandwich})). 

%

\section{Scale separation with fixed $\eta$, $B$ and varying $N$} \label{scale_separation}

In this section we investigate how the scale separation, i.e. ratio $B$ to $N$ affects the accuracy of reconstruction of the  Jacobian and velocity. We use two test initial conditions as before with Gaussian window function $\psi^{(6)}(x)$, $\eta=0.1$,  $B=500$ and  $N=1000$, $2000$, $5000$ and $10,\!000$. 

\begin{figure}[h] \centerline{
\includegraphics[height=1.5in,angle=0]{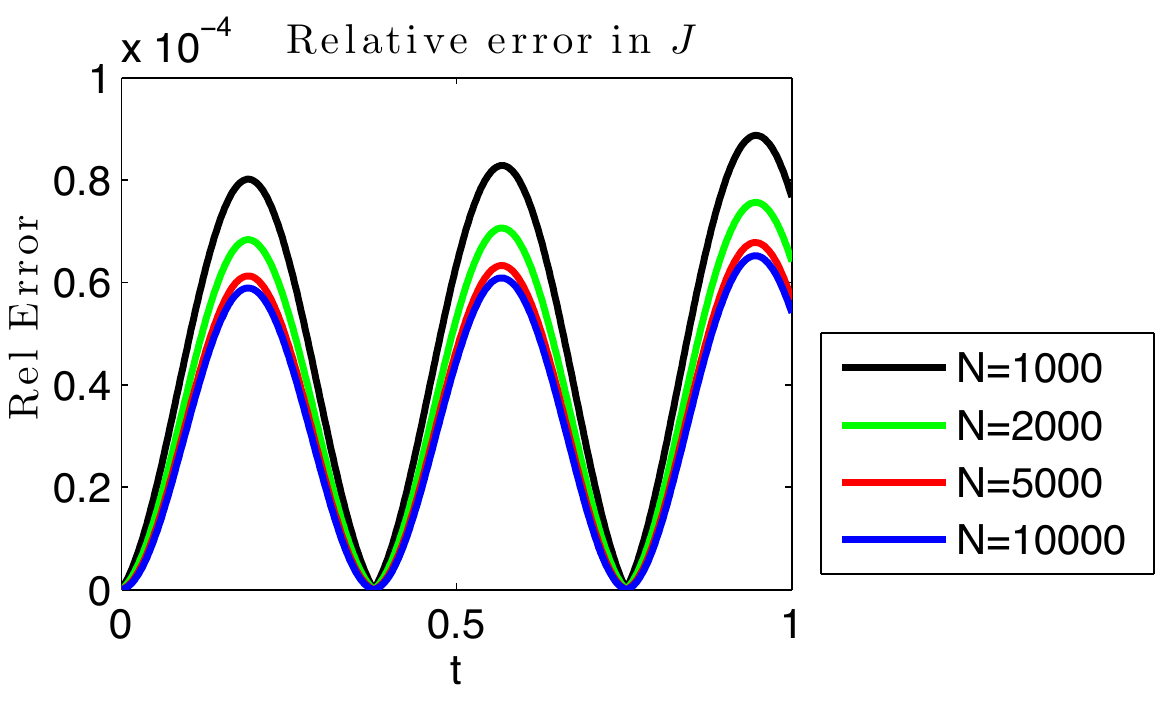}
\includegraphics[height=1.5in,angle=0]{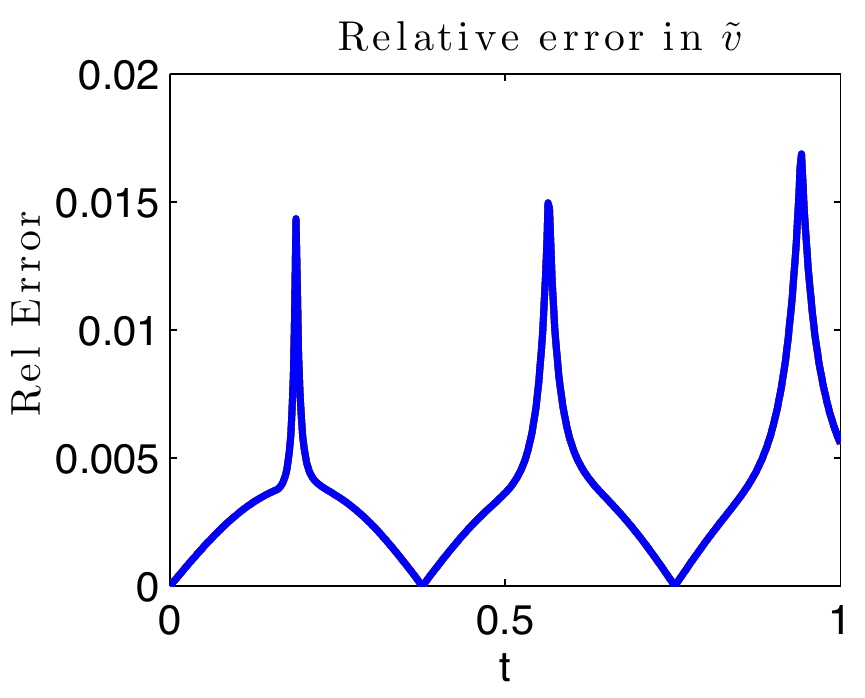}
} \caption{Effect of the scale separation on the reconstruction of the Jacobian (left panel) and velocity (right panel) in the first test case with   $\eta=0.1$, $\psi^{(6)}(x)$, $B=500$ and $N=1000$, $2000$, $5000$ and $10,\!000$.}
\label{W9a}
\end{figure}

\begin{figure}[h] \centerline{
\includegraphics[height=1.5in,angle=0]{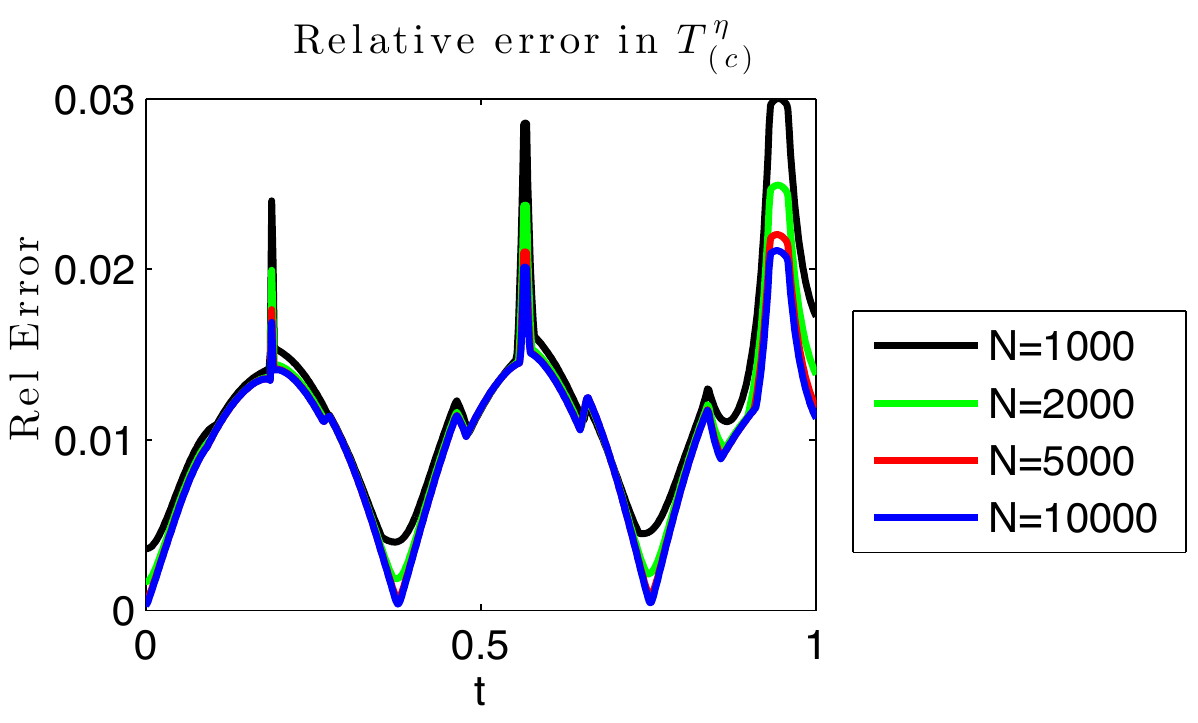}
\hspace{20pt}
\includegraphics[height=1.5in,angle=0]{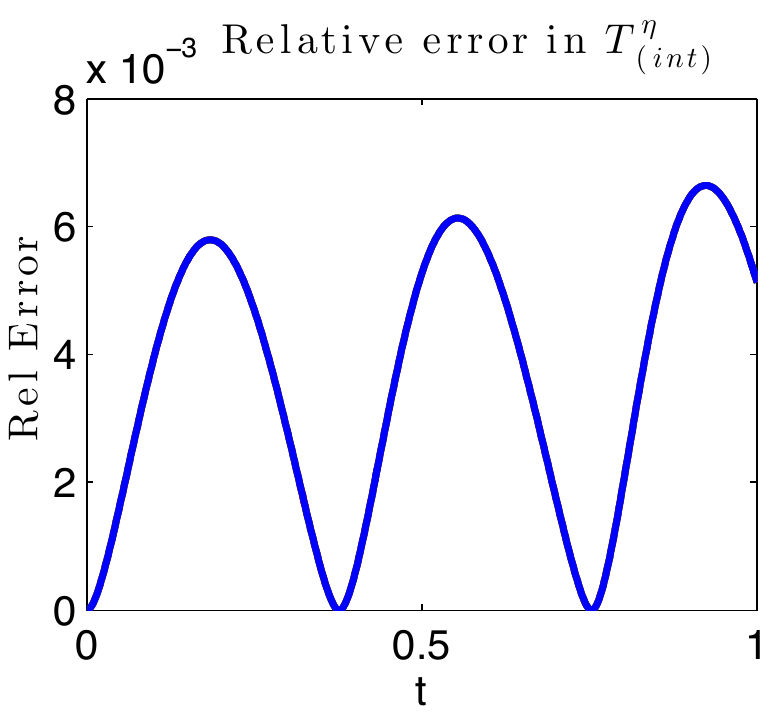}
} \caption{Effect of the scale separation on the error in approximation of the convective  stress (left panel) and  
interaction stress (right panel) for the first test case with $\eta=0.1$, $\psi^{(6)}(x)$, $B=500$ and $N=1000$, $2000$, $5000$ and $10,\!000$.}
\label{W9}
\end{figure}

In the first test case, results of which are presented in Figs. \ref{W9a} and \ref{W9}, we observe that the error in reconstruction of the Jacobian (shown in the left panel of Fig. \ref{W9a}) increases as $N$ increases, while the error in velocity reconstruction (right panel of Fig. \ref{W9a}) does not depend on $N$. Both errors oscillate in time and their oscillatory dynamics is related to the total computed energy oscillations depicted in Fig. \ref{W8b} in the left panel, i.e. when energy starts deviating from its initial (exact) value, error in the Jacobian reconstruction starts to increase and when the energy starts returning to its initial value, the error in the Jacobian starts decreasing.  The error in the velocity approximation does not essentially depend on $N$, oscillates as well and does not exceed $2\%$ during the simulation time. Even though the reconstruction of the Jacobian becomes worse as the scale separation increases, the approximation of the convective stress (shown in the left panel of Fig. \ref{W9}) gets better as $N$ increases. At all times, the error in the convective stress does not exceed $3\%$. The error in the interaction stress does not depend on $N$ as can be seen from the right panel of Fig. \ref{W9}.  

In the second test case reconstruction of both Jacobian and velocity depends on $N$ in a non-monotonic manner. The intervals where the error with $N=10,000$ is smaller and larger alternate in phase with oscillations of the total computed energy.
%
%
%
While the error in the Jacobian approximation is at most $0.2\%$, the error in the velocity approximation is rather large, especially at later times: it is below $5\%$ until $t=0.7 $ and then increases to $15-25\%$ ($15\%$ for $10,000$).
The error in the approximation of the convective stress is shown in the left panel of Fig. \ref{W9}. It is not monotonic in $N$, and it does not exceed $6\%$. The error in the interaction stress does not depend on $N$ and stays below $2.5\%$.

\begin{figure}[h] \centerline{
\includegraphics[height=1.5in,angle=0]{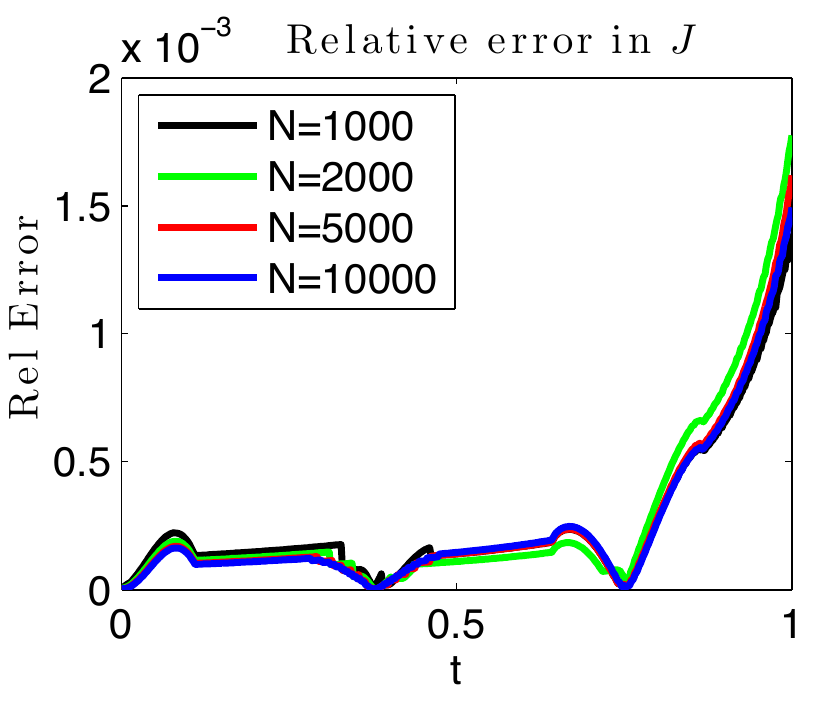}
\hspace{20pt}
\includegraphics[height=1.5in,angle=0]{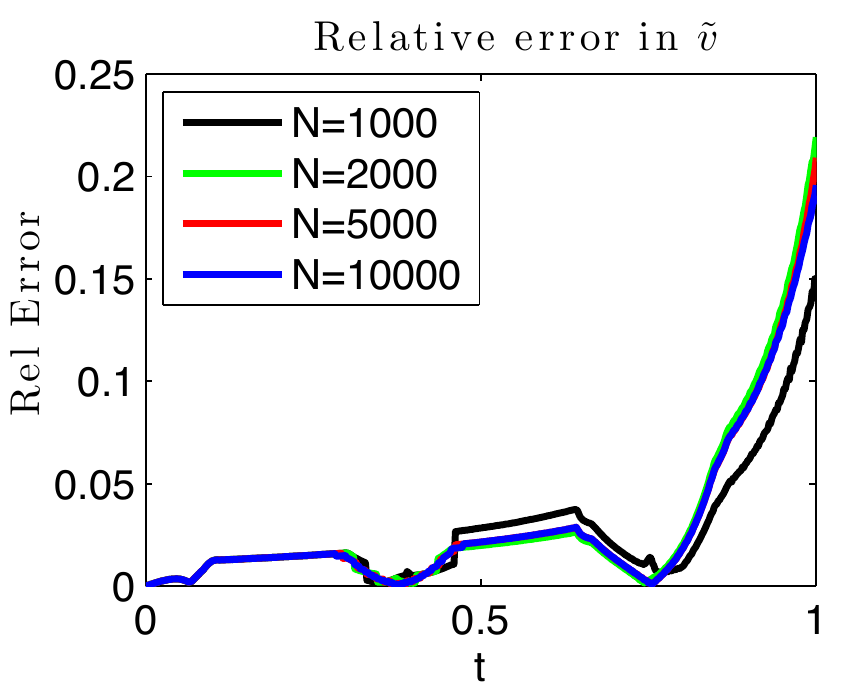}
} \caption{Effect of the scale separation on the reconstruction of the Jacobian (left panel) and velocity (right panel) in the second test case with   $\eta=0.1$, $\psi^{(6)}(x)$, $B=500$ and $N=1000$, $2000$, $5000$ and $10,\!000$.}
\label{W11Jac}
\end{figure}

\begin{figure}[h] \centerline{
\includegraphics[height=1.5in,angle=0]{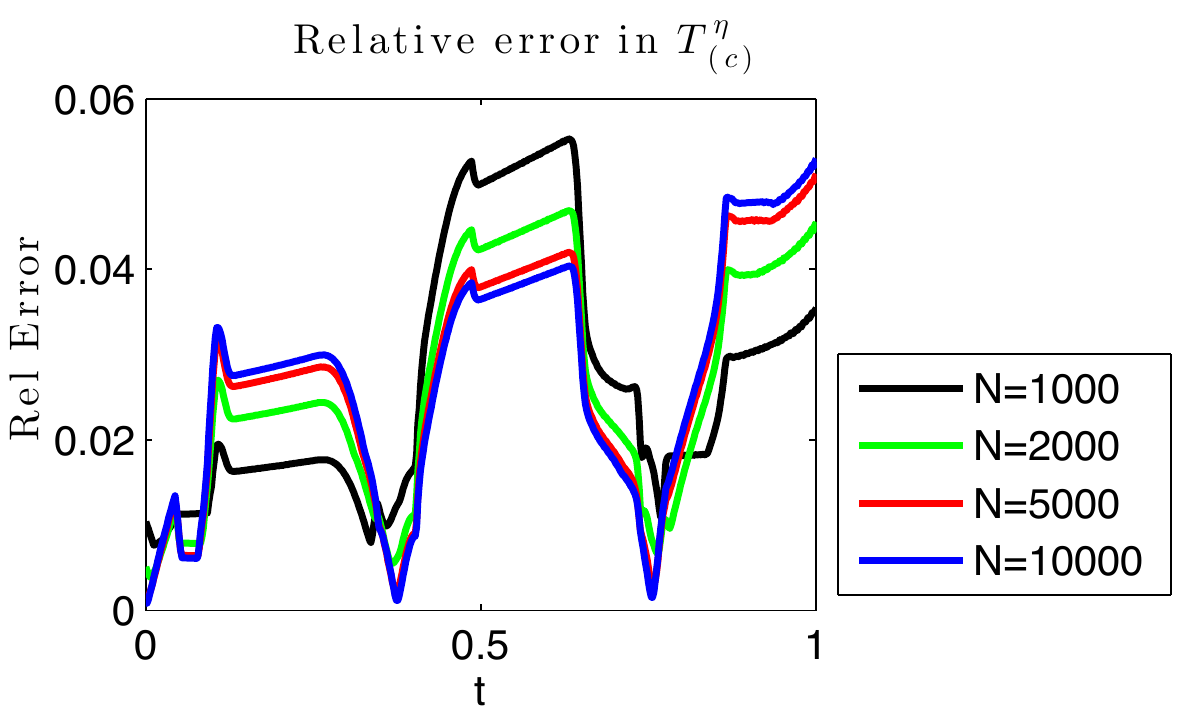}
\includegraphics[height=1.5in,angle=0]{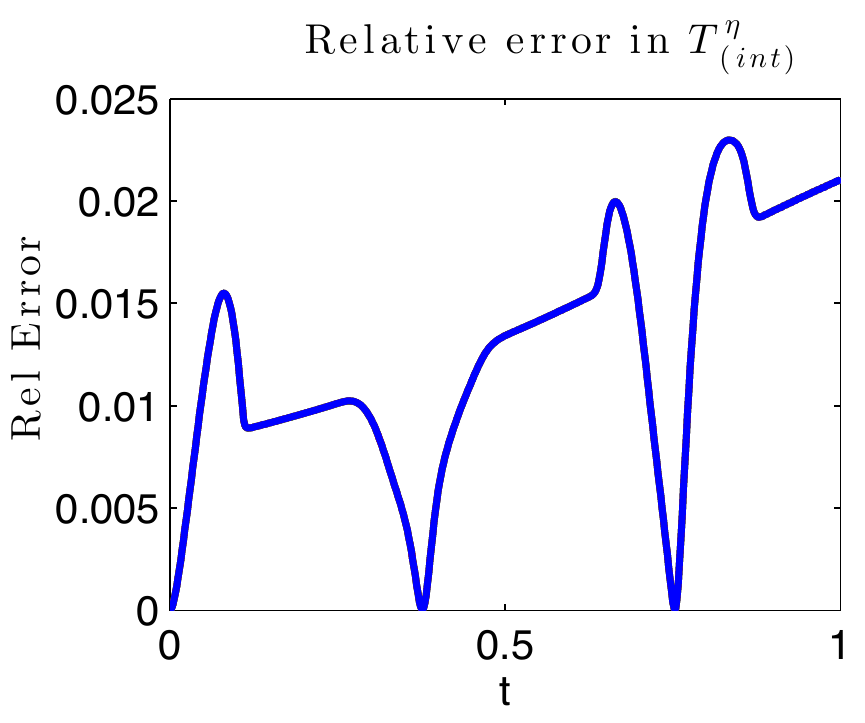}
} \caption{Effect of the scale separation on the error in approximation of the convective  stress (left panel) and  
interaction stress (right panel) for the second test case with $\eta=0.1$, $\psi^{(6)}(x)$, $B=500$ and $N=1000$, $2000$, $5000$ and $10,\!000$.}
\label{W11}
\end{figure}

%
\section{Error estimates}\label{error_estimates}
\subsection{Estimates for filtered regularization methods} 

In practice, the right hand side of (\ref{e1}) is known imprecisely, so instead of the exact $\fbb$, one has an approximate vector $\fbb^\delta$. The computed regularized solution is thus
\begin{equation}
\label{e10}
\bx^{\alpha, \delta}=\sum_{j=1}^D b_j^\delta \frac{\phi(\sigma_j, \alpha)}{\sigma_j}\hat\bxi_j.
\end{equation}

Our goal is to estimate the error $\parallel \bx-\bx^{\alpha, \delta}\parallel_p$, in some vector $p$-norm. Usually $p=2$, but here we assume $p\in [1, \infty)$. By triangle inequality,
\begin{eqnarray}
\label{triangle}
\parallel \bx-\bx^{\alpha, \delta}\parallel_p & \leq &
\left|\left|\sum_{j=1}^D b_j \frac{1-\phi(\sigma_j, \alpha)}{\sigma_j}\hat\bxi_j\right|\right|_p+
\left|\left|
\sum_{j=1}^D (b_j-b_j^\delta)\frac{\phi(\sigma_j, \alpha)}{\sigma_j}\hat\bxi_j\right|\right|_p\\
& \leq & C(\hat \bxi, p) \left(
\left(\sum_{j=1}^D |b_j|^p \frac{|1-\phi(\sigma_j, \alpha)|^p}{\sigma_j^p}\right)^{\frac{1}{p}}+
\left(\sum_{j=1}^D |b_j-b_j^\delta|^p\frac{|\phi(\sigma_j, \alpha|^p}{\sigma_j^p}\right)^{\frac{1}{p}}
\right).\nonumber
\end{eqnarray}
The constant $C(\hat\bxi, p)$ depends only on $p$ and the components of the singular vectors $\hat\bxi_j$.

Writing $|b_j|/\sigma_j=|x_j|$, we see that the first term on the very right of (\ref{triangle}) is bounded
by
\begin{equation}
\label{e11}
C(\hat \bxi, p)
\max_{j}|x_j|\left(\sum_{j=1}^D|1-\phi(\sigma_j, \alpha)|^p\right)^{\frac{1}{p}}\leq
C_1(\hat\bxi, p) \parallel \bx\parallel_\infty
\left(\int_0^{D+1} |1-\phi(f(t), \alpha)|^p dt\right)^{\frac{1}{p}},
\end{equation}
where we introduced a function $f(t): [0, \infty)\to (0, 1]$ that interpolates between singular values:
 $$
f(j)=\sigma_j, \;\;\;f(0)=1, \;\;\;\lim_{t\to\infty} f(t)=0.
$$
The function $f$ is chosen to be continuous, non-negative, and strictly decreasing. To see that the integral is larger than the corresponding sum, note that the sum is the left-endpoint Riemann sum for the integral, and
the function under the integral is increasing.

Similarly,
\begin{equation}
\label{e12}
C(\hat \bxi, p)
\left(\sum_{j=1}^D |b_j-b_j^\delta|^p\frac{|\phi(\sigma_j, \alpha|^p}{\sigma_j^p}\right)^{\frac{1}{p}}
\leq
C_1(\hat\bxi, p) \parallel \fbb-\fbb^\delta \parallel_\infty
\left(\int_0^{D+1} \frac{|\phi(f(t), \alpha)|^p}{f(t)^p} dt\right)^{\frac{1}{p}}.
\end{equation}

Combining (\ref{triangle})-(\ref{e12}) we have
\begin{eqnarray}
\label{e13}
\parallel \bx-\bx^{\alpha, \delta}\parallel_p & \leq &
C_1(\hat\bxi, p) \parallel \bx\parallel_\infty
\parallel 1-\phi(f(t), \alpha)\parallel_{L^p(0, D+1)}\\
&+ &
C_1(\hat\bxi, p) \parallel \fbb-\fbb^\delta \parallel_\infty
\parallel \phi(f(t), \alpha) f^{-1}(t)\parallel_{L^p(0, D+1)}.\nonumber
\end{eqnarray}

By definition of $\phi$, the first term can be made arbitrarily small by choosing $\alpha$ small enough. In the second term, as $\alpha\to 0$, the norm of $\phi(f(t), \alpha) f^{-1}(t)$ typically increases. To control the second term, we need $\fbb-\fbb^\delta$ to be small. This is typical of the error estimates available in the literature. Our inequalities differ from the standard ones because we use a $p$-norm for the error, and $\infty$-norms
for $\bx$ and $\fbb-\fbb^\delta$. The standard estimates use $2$-norms
of $\bx-\bx^{\alpha, \delta}$, $\bx$, $\fbb-\fbb^\delta$, and what is essentially the $\infty$-norm for the
$\phi$- and $f$-dependent terms.
Depending on the actual $\phi$ and $f(t)$, our approach can yield tighter bounds. Improvement occurs if, loosely speaking, the integrals involving $\phi, f$ in (\ref{e11}), (\ref{e12}) are smaller than maximal values of the integrands.

Finally, we note that using H\" older inequality with exponents $q, q^\prime, q^{-1}+(q^\prime)^{-1}=1$ in the right hand side of (\ref{triangle}) results in the estimates
\begin{eqnarray}
\label{e14}
\parallel \bx-\bx^{\alpha, \delta}\parallel_p & \leq &
C_2(\hat\bxi, p, q) \parallel \bx\parallel_{pq}
\parallel 1-\phi(f(t), \alpha)\parallel_{L^{pq^\prime}(0, D+1)}\\
&+ &
C_2(\hat\bxi, p, q) \parallel \fbb-\fbb^\delta \parallel_{pq}
\parallel \phi(f(t), \alpha) f^{-1}(t)\parallel_{L^{pq^\prime}(0, D+1)}.\nonumber
\end{eqnarray}


\subsection{Error in the interaction stress approximation}
The purpose of this section is to estimate the difference between the exact integral representation of the interaction stress $\bT^\eta_{(int)}$
in (\ref{c2-int}), and its closed form approximation
\begin{equation}
\label{int-closed}
\overline{\bT}^\eta_{(int)}(t, \bx)=
\frac{1}{|\Omega|^2}\int \psi_\eta(\bx-\bR) 
\left(\int U^\prime(|\brho|)\frac{\brho\otimes \brho}{|\brho|}
Q_\eta[\overline{\rho}^\eta](t, \bR+\frac{\vep}{2}\brho)Q_\eta[\overline{\rho}^\eta](t, \bR-\frac{\vep}{2}\brho) d\brho
\right)\; d \bR.
\end{equation}
Since estimates will be local in time, we will suppress the dependence on $t$ in the remainder of this section. Define the error
$$
E(\bx)=\bT^\eta_{(int)}(\bx)-\overline{\bT}^\eta_{(int)}(\bx).
$$
Next, introduce the abbreviated notation 
$$
J^+=J(\bR+\frac{\vep}{2}\brho), \qquad J^-=J(\bR-\frac{\vep}{2}\brho),
$$
and 
$$
Q_\eta[\overline{\rho}^\eta]^+ =Q_\eta[\overline{\rho}^\eta](\bR+\frac{\vep}{2}\brho),  \qquad
Q_\eta[\overline{\rho}^\eta]^- =Q_\eta[\overline{\rho}^\eta](\bR-\frac{\vep}{2}\brho),
$$
and
denote
\begin{equation}
\label{Fi}
\Phi(\brho)=U^\prime(|\brho|)\frac{\brho\otimes\brho}{|\brho|}.
\end{equation}
This function is smooth and can be assumed compactly supported on a shell $D=\{\brho: c_1\leq |\brho|\leq c_2\}$ where $c_1>0$.  
With these notations, using an elementary identity
$$
a_1 a_2- b_1 b_2=a_1(a_2-b_2)+a_2(a_1-b_1)-(a_1-b_1)(a_2-b_2)
$$ 
we have 
\begin{eqnarray}
\label{split}
J^+J^- - Q_\eta[\overline{\rho}^\eta]^+ Q_\eta[\overline{\rho}^\eta]^- 
& = &
J^+(J^- -Q_\eta[\overline{\rho}^\eta]^-)+
(J^+ -Q_\eta[\overline{\rho}^\eta]^+)Q_\eta[\overline{\rho}^\eta]^- \\
& = &
J^+(J^--Q_\eta[\overline{\rho}^\eta]^-)+J^-(J^+-Q_\eta[\overline{\rho}^\eta]^+)\nonumber \\
& - &
(J^--Q_\eta[\overline{\rho}^\eta]^-)(J^+-Q_\eta[\overline{\rho}^\eta]^+).
\nonumber
\end{eqnarray}
Now
\begin{eqnarray*}
|E(\bx)|& \leq  & |\Omega|^{-2} \sup |\psi_\eta|
\int\int |\Phi(\brho)||J^+ J^- - Q_\eta[\overline{\rho}^\eta]^+ Q_\eta[\overline{\rho}^\eta]^-|(\bR, \brho) d\brho\;d\bR\\
& \leq &
|\Omega|^{-2} \sup |\psi_\eta|
\sup_D |\Phi|\\
& &
 \int\int 
 \biggl(
|J^+| |J^- - Q_\eta[\overline{\rho}^\eta]^-|+
|J^-| |J^+ - Q_\eta[\overline{\rho}^\eta]^+|+
|J^+ - Q_\eta[\overline{\rho}^\eta]^+| |J^- - Q_\eta[\overline{\rho}^\eta]^-| \biggr)d\brho\;d\bR.
\end{eqnarray*}
Changing variables in the last integral to 
$$
\by_1=\bR+\frac{\vep}{2}\brho, \qquad \by_2=\bR-\frac{\vep}{2}\brho,
$$
and observing that the Jacobian of this transformation is $\vep^{-d}$ and that the quantities marked by $^+$ (respectively by $^-$) depend only on
$\by_1$ (respectively on $\by_2$),
we find 
$$
|E(\bx)|\leq \vep^{-d}
|\Omega|^{-2} C( \psi_\eta, \Phi)
\left[
2 \int_\Omega |J(\by_1)|d\by_1
\int_\Omega |J- Q_\eta[\overline{\rho}^\eta]|(\by_2) d\by_2+
\left(\int_\Omega |J- Q_\eta[\overline{\rho}^\eta]|(\by)d\by\right)^2
\right].
$$
Suppose now that $J$ and $Q_\eta[\overline{\rho}^\eta]$ are given by their discretizations on the fine mesh. Thus we can assume that
they are piecewise constant functions having values $J_j$, $Q_\eta[\overline{\rho}^\eta]_j$ on the sets $S_j\subset \Omega, j=1, 2, \ldots, N$ of measure
$|\Omega|/N$. In this way, $J$ and $Q_\eta[\overline{\rho}^\eta]$ can be identified with, respectively,  the vectors $\bJ=(J_1, J_2, \ldots, J_N)^T$ and
$\bQ=(Q_1, Q_2, \ldots, Q_N)^T$.

Suppose that there exists a constant $M$ such that
\begin{equation}
\label{M}
J\leq M.
\end{equation}
With this, 
\begin{eqnarray}
\label{E}
|E(\bx)|& \leq &
\vep^{-d} |\Omega|^{-2}
C(\psi_\eta, \Phi) 
\left(\frac{|\Omega|}{N}
\right)^2
\left(
2M \parallel \bJ-\bQ\parallel_1+\parallel \bJ-\bQ \parallel_1^2
\right)\\
& = & 
C(\psi_\eta, \Phi)
\left(
2M \parallel \bJ-\bQ\parallel_1+\parallel \bJ-\bQ \parallel_1^2
\right).\nonumber
\end{eqnarray}
The last equality holds since $N=\vep^{-1/d}$. The norms are vector 1-norms that can be estimated using
(\ref{e14}) with $\bx=\bJ$, $\bx^{\alpha, \delta}=\bQ$, and $\fbb$ representing a discretization of $\overline{\rho}^\eta$.

The results of this section can be summarized in the
\begin{theorem}
Suppose that \newline
\noindent
(i) $J(t, \bx)$ satisfies (\ref{M}) uniformly in $t$;\newline
\noindent
(ii)  $\Phi$ defined in (\ref{Fi}) is bounded;\newline
\noindent
(iii) $J(t, \bx)=\sum_{j=1}^N J_j(t, \bx) \chi_j(\bx)$, $Q_\eta[\overline{\rho}^\eta]=\sum_{j=1}^N Q_j(t, \bx) \chi_j(\bx)$, where
$\chi_j$ are characteristic functions of sets $S_j$ such that $\cup_{j=1}^N=\Omega$, $S_j\cap S_k =\emptyset$ if $j\ne k$, and
$|S_j|=N^{-1}|\Omega|$. 
Then the error 
$$
E=\bT^\eta_{(int)}-\overline{\bT}^\eta_{(int)}
$$
satisfies
$$
|E(t, \bx)|
\leq
\sup |\psi_\eta| \sup |\Phi|
\left(
2M \parallel \bJ-\bQ\parallel_1+\parallel \bJ-\bQ \parallel_1^2
\right),
$$
where $\bJ=(J_1, J_2, \ldots, J_N)^T$, $\bQ=(Q_1, Q_2,\ldots, Q_N)^T$.
\end{theorem}

The estimates for the error in $\bT^\eta_{(c)}$ can be derived similarly, but would require more technical work because of the triple product structure of the integrand. It is also worth noting that, while a pointwise bound on $J$ can be reasonably expected, similar bounds on the velocity $\tilde v$ would blow up as $\vep\to 0$. Consequently, estimating of the error $\bT^\eta_{(c)}-\overline{\bT}^\eta_{(c)}$ is left to future work.

\section{Conclusions} \label{conclusions}

We study the numerical performance of the regularized deconvolution closure  introduced in \cite{Panchenko_Barannyk_Gilbert_2011, Panchenko_Barannyk_Cooper}. 
The closure method consists of the following.  The average density and linear momentum are written as convolutions acting on respective fine scale functions: $J$ and $J\tilde\bv$, where $J$ is the Jacobian of the inverse deformation map, and $\tilde \bv$ is a particle velocity interpolant. These functions are approximately recovered by applying a regularized deconvolution to the averages. To construct the deconvolution operator, we use the theory of ill-posed problems.
Closure is obtained by using these deconvolution approximations in the exact flux equations. This gives  constitutive equations that express stress in terms of the average density and velocity. 
The exact stress is thus approximated by  a sum of terms that have the ``convolution sandwich"  structure: they combine the  convolution  operator,  a nonlinear composition or a product type operator, and the  deconvolution operator. The resulting constitutive equations are nonlinear and nonlocal.

The approximation quality depends on a choice of the window function $\psi$ used to define averages, magnitude of scale separation, and values of the resolution and regularization parameters.
Because of the nonlinearity of the problem, the error estimates tend to be too pessimistic. Therefore,  we conduct numerical experiments to determine the dependence of the error on the above parameters. 
Since the Fourier spectrum of velocity seems to have a strong effect on the error, 
we consider two sets of initial conditions. In the first test case, the initial velocity  is a low frequency mode sine function, while in the  second test the initial velocity has full Fourier spectrum. The initial positions in both cases are equally spaced. 

We study  window functions of different smoothness, varying from piecewise continuous to infinitely smooth.
Among these functions, the Gaussian provides the best overall performance 
despite the fact that the corresponding integral deconvolution problem  has the highest degree of ill-posedness.
Numerical deconvolution amounts to solving an ill-conditioned linear system. We use  a truncated SVD method with an additional spectral filtering of the right hand side. Filtering helps to reduce the effect of error that is present in every standard numerical SVD routine. 

The choice of the resolution parameter $\eta$ affects the size of the averaging window and the amount of high frequency filtering in the computed averages. Larger values of $\eta$ produce smoother and smaller averages, thus causing the reconstruction to deteriorate. This tendency is counteracted by the presence of the convolution operator in the stress equations. We find that the overall error in the stress approximation tends to decrease with increasing $\eta$. Therefore, it is not necessary to have very good reconstruction of the Jacobian and velocity to have good approximations of the stresses. This ``self-correcting" property is a noteworthy feature of the deconvolution closure.
Another method to increase scale separation is to vary the number of particles while keeping $\eta$ fixed. The results in this case are less clear-cut compared to the case of increasing $\eta$. However, at times when the computed total energy is close to its exact value, the error in the stress decreases with increasing scale separation.

The deconvolution error estimates derived in the paper are applicable to general SVD-based filtered regularization methods (see e.g. \cite{Kirsch_1996}). 
We also obtain error estimates for the interaction stress (the part of the total stress induced by interparticle forces). We believe that similar estimates can be also obtained for the remaining convective stress, but such estimates will be developed elsewhere.


\appendix

\section{Window functions} 

A window function $\psi$ is  chosen to define a mesoscale average. This function has to satisfy several conditions: be nonnegative, fast decreasing, compactly supported (we also consider non-compactly supported functions like Gaussian), continuous and differentiable almost everywhere in the interior of its domain and  $\int_{\infty}^\infty \psi(x)=1$. We use functions $\psi^{(i)}(x)$, $i=1,\ldots,6$ of different order of smoothness starting from the characteristic function $\psi^{(1)}(x)$ that is discontinuous at $x=\pm \frac L2$ up to infinitely many times differentiable Gaussian $\psi^{(6)}(x)$. The window functions are defined in (\ref{psi1})-(\ref{psi6}) and plotted in Fig. \ref{W1}.

\begin{equation}\label{psi1}
\hspace*{-75pt}
\psi^{(1)}(x)=
\left\{
\begin{array}{l}
\displaystyle 1/L, \quad \mbox{if} \quad |x|\leq  L/2, \\[5pt]
0, \quad \hspace{4pt} \mbox{otherwise};
\end{array}
\right.
\end{equation}

\begin{equation}
\hspace*{35pt}
\psi^{(2)}(x)=
\left\{
\begin{array}{l}
\displaystyle {1}/({2L}), \quad \mbox{if} \quad |x|\leq  L/2, \\[5pt]
\displaystyle -2/L\bigl(x-{3L}/{2}\bigr), \quad \mbox{if} \hspace{12pt} \quad  L/2< x\leq {3L}/{2}, \\[5pt]
\displaystyle -2/L\left(x+{3L}/{2}\right), \quad  \mbox{if}  \quad -{3L}/{2} \leq  x<-{L}/{2}, \\[5pt]
0, \quad \mbox{otherwise};
\end{array}
\right.
\label{psi2}
\end{equation}

\begin{equation}
\psi^{(3)}(x)=
\left\{
\begin{array}{l}
\displaystyle-{4}/{L^2}(x- L/2), \quad \mbox{if} \quad 0\leq x\leq  L/2, \\[5pt]
\displaystyle{4}/{L^2}(x+ L/2), \quad \mbox{if} \quad - L/2\leq x< 0, \\[5pt]
0, \quad \mbox{otherwise};
\end{array}
\right.
\label{psi3}
\end{equation}

\begin{equation}
\psi^{(4)}(x)=
\left\{
\begin{array}{l}
\displaystyle-{6}/{L^3}\biggl(x^2-{L^2}/{4}\biggr), \quad \mbox{if} \quad |x|< L/2, \\[5pt]
0, \quad \mbox{otherwise};
\end{array}
\right.
\label{psi4}
 \end{equation}

\begin{equation}
\psi^{(5)}(x)=
\left\{
\begin{array}{l}
\displaystyle {30}/{L^5}\left(x^2-{L^2}/{4}\right)^2, \quad \mbox{if} \quad |x|\leq  L/2, \\[5pt]
0, \quad \mbox{otherwise};
\end{array}
\right.
\label{psi5}
\end{equation}

\begin{equation}
\hspace*{-80pt}
\psi^{(6)}(x)=
\frac{6}{L\sqrt{2\pi}}\exp\biggl(-\frac{18x^2}{L^2}\biggr).
\label{psi6}
\end{equation}

\section{Lennard-Jones potential}

The dynamics of particles considered in this paper is  governed by Lennard-Jones potential defined as
\begin{equation}\label{LJ_def}
U(\xi) = 4\epsilon\left[\left(\frac{\sigma}{\xi}\right)^{12}-\left(\frac{\sigma}{\xi}\right)^{6}\right],
\end{equation}
with the potential well depth $\epsilon=0.025$. The same potential but with $\epsilon=0.25$ was used in \cite{Panchenko_Barannyk_Cooper}. The magnitude of   $\epsilon$  defines how strong interaction between particles is. The potential is zero at the distance given by  $\sigma$ and reaches its minimum at the distance $h=2^{1/6}\sigma$ at which particles are in equilibrium. For smaller distances $\xi<h$, the potential is repulsive whereas for $\xi>h$ it is mildly attractive. When the distance $\xi>2.5h$, the force is very small and we set it to zero to speed up computations. This truncation of the potential tail typically takes into account  $3$ particles on each side from a current particle. Truncating at larger distances slightly decreases deviations of the total computed energy from its exact value (at $t=0$) and as a result slightly decreases the error in the approximation of the stresses, more so in the interaction stress.

%
%

%

\bibliography{references_study_deconvolution}

\end{document}